\documentclass[a4paper,11pt]{article}
\usepackage{a4wide}
\usepackage{graphicx}
\usepackage{amsmath}
\usepackage{xcolor}
\def\sech{\mathop{\rm sech}\nolimits}

\title{Nematic dispersive shock waves from nonlocal to local}
\author{Saleh Baqer,\\
Department of Mathematics, Faculty of Science, \\
Kuwait University, Kuwait City 13060, Kuwait \and
Dimitrios J. Frantzeskakis,\\
Department of Physics, \\
National and Kapodistrian University of Athens, 157 84 Athens, Greece. \and
 Theodoros Horikis,\\
Department of Mathematics, \\
University of Ioannina, Ioannina 451 10, Greece \and
C\^{o}me Houdeville, \\
Ecole Nationale Sup\'erieure de Techniques Avanc\'ees,\\
828 Boulevard des Mar\'echaux, 91120 Palaiseau, France. \and
Timothy R. Marchant,\\
School of Mathematics and Applied Statistics,\\
University of Wollongong,\\
Northfields Avenue, Wollongong, New South Wales, Australia, 2522.\and
Noel F. Smyth, \\
School of Mathematics, University of Edinburgh,\\
Edinburgh, Scotland, EH9 3FD, U.K. \\ and \\
School of Mathematics and Applied Statistics,\\
University of Wollongong,\\
Northfields Avenue, Wollongong, New South Wales, Australia, 2522.}
\date{}

\begin{document}

\maketitle

\begin{abstract}
The structure of optical dispersive shock waves in
nematic liquid crystals is investigated as the power of the optical beam is varied,
with six regimes identified, which complements previous work pertinent to
low power beams only. It is found that the dispersive shock wave structure depends
critically on the input beam power. In addition, it is known that nematic dispersive
shock waves are resonant and the structure of this resonant is also critically dependent
on the beam power. Whitham modulation theory is used to find solutions for the six regimes with
the existence intervals for each identified. These dispersive shock wave solutions
are compared with full numerical solutions of the nematic equations and excellent
agreement is found.
\end{abstract}

\section{Introduction}

Nematic liquid crystals form an ideal medium to study nonlinear optics due to their ``huge''
nonlinearity, which is orders of magnitude larger than that of optical fibres, so that nonlinear
effects can be observed over millimetre distances \cite{khoo,PR,Wiley,conti2}. In particular, the
refractive index of nematic liquid crystals increases with optical intensity, so that they form a
focussing medium. When a light beam propagates through a nematic liquid crystal, the electric field
of the electromagnetic wave induces dipoles in the nematic molecules, which then rotate, changing
the refractive index. In addition, nematic liquid crystals have a non-local response to an optical
beam in that the elastic response of the nematic extends far beyond the optical forcing
\cite{conti2}. An optical beam propagating through a nematic medium can then form its own
waveguide, resulting in a self-guided beam, an optical solitary wave, termed a ``nematicon''
\cite{PR,Wiley,physdreview,assantokhoo}, which was first experimentally generated and observed in 2000 \cite{assantokhoo}. Since this first observation, nematicons, and related solitary-type waves, such as optical vortices, have become a theme of intense
experimental and theoretical research effort,
driven both by interest in the nonlinear optics of nematic liquid crystals
and also by their potential applications in optical devices \cite{steer,spatial,readdress,molding};
see Refs.~\cite{PR,Wiley,physdreview,minzonireview} for general reviews on the nonlinear optics
of nematic liquid crystals.

Solitary waves 
are generic wave forms for nonlinear dispersive wave equations \cite{whitham},
first observed and studied in the context of water waves \cite{whitham,russell} and
fluid dynamics \cite{whitham}, but are widespread in nature
arising, e.g., in solid mechanics \cite{karima}, biology \cite{davydov}, ecology \cite{tlidi}, and the above mentioned context of
nonlinear optics \cite{PR,kivshar,ablowitz}, for instance. One of the
appealing features of solitary waves, in addition to their widespread occurrence in nature,
is that they are 
localized waves with steady profiles,
which makes them
easier to study theoretically. In addition to this, many generic nonlinear dispersive wave equations, such as the Korteweg-de Vries (KdV), nonlinear Schr\"odinger (NLS) and Sine-Gordon equations, are completely integrable systems via
the Inverse Scattering Transform method \cite{whitham,ablowitz}.
Thus, a general initial condition for these equations will form a finite number
of solitary waves, plus dispersive radiation. In addition, solitary wave solutions of integrable
nonlinear dispersive wave equations, solitons, interact ``elastically'',
i.e., they emerge unscathed out of the interaction without any change in their form; 
hence, due to this particle-like behaviour, solitary waves are termed solitons for such equations.

Another generic wave form supported by nonlinear dispersive wave equations are the
dispersive shock waves (DSWs), also termed undular bores; these structures are as widespread
in nature as solitary waves, with well known examples being tidal bores and tsunamis
\cite{elreview}.  In contrast to a solitary wave, a DSW is a non-steady wave form which
continuously expands. A DSW is a dispersive regularization
of a discontinuity and is a modulated periodic wavetrain with solitary waves at one edge
and linear, dispersive waves at the other--- see \cite{elreview} for a general review of DSWs.
Since DSWs are non-steady waveforms, their study is more difficult than that for solitary waves.
The development of DSW solutions of nonlinear dispersive wave equations relies chiefly on
Whitham modulation theory \cite{whitham,mod1,modproc,modbenjamin}, which is a version of the
asymptotic method of multiple scales that is used to analyse slowly varying periodic wavetrains.
Whitham modulation equations are a system of partial differential equations which govern the
parameters of a slowly varying wavetrain, such as its amplitude, wavenumber, frequency and mean
height. If this system is hyperbolic, then the underlying wavetrain is modulationally stable, while
if it is elliptic, the wavetrain is unstable \cite{whitham}.  A major achievement of Whitham
modulation theory was the development of the modulation equations for the KdV equation
\cite{whitham,modproc}. These modulation equations form a hyperbolic system, so that the cnoidal
wave solution of the KdV equation is modulationally stable. It was subsequently realized that a
simple wave solution of the KdV modulation equations is a DSW \cite{gur}, even though the initial
condition is a step, which is not slowly varying.  This DSW solution is in excellent agreement with
numerical solutions of the KdV equation \cite{bengt}.
The key to the determination of the simple wave DSW solution is the ability to set the modulation
equations in Riemann invariant form. If the nonlinear dispersive wave equation governing the DSW is
integrable, 
then its Whitham modulation equations can be automatically set in Riemann invariant form \cite{flash}, so that the DSW solution can easily be found.

As mentioned above, the standard DSW form, termed of KdV type \cite{elreview}, is a modulated
periodic wave with solitary waves at one edge and linear dispersive waves at the other.
A non-standard DSW type is a resonant DSW \cite{patkdv,pat}, for which the waves of the DSW are in
resonance with (linear) dispersive waves, resulting in a resonant wavetrain being emitted from the
DSW. Resonant DSWs also occur for the KdV equation with next higher-order dispersion, i.e.,
fifth-order dispersion, namely for
the Kawahara equation \cite{kaw}, and the NLS equation with next order,
third-order dispersion \cite{trillores,trilloresfour,trilloresnature,trilloreslossbore}.
If the emitted resonant wavetrain is of large enough amplitude,
the KdV-type DSW structure can be destroyed;
this results in the so-called traveling dispersive shock wave (TDSW) regime \cite{patkdv,pat},
consisting of a resonant wavetrain with negative polarity solitary wave, which is the
remnant of the DSW, linking this to the level behind \cite{patkdv}--- see Figure \ref{f:types}(d)
below for an example of such a TDSW. A nematic liquid crystal is a focusing medium and, thus,
optical waves are modulationally unstable; as a result,
an optical DSW is not supported. However, the addition of azo dyes to the nematic medium
changes its response so that it becomes defocusing \cite{gaetanodark}; in this case, nematics
can support DSWs \cite{nembore,nemgennady,saleh}. A nematic DSW is an example of a resonant DSW
\cite{nembore,nemgennady,saleh}. In these works,
the nematic DSW was studied in the highly nonlocal limit, for which the nematic elastic response
extends far beyond the light beam, with the nematic DSW generated by a step jump in the optical intensity.
While the nematic equations are of NLS-type \cite{PR}, in the highly nonlocal limit the
nematic bore is of KdV-type and is well described by the DSW solution of the KdV equation.  The
nematic DSW structure is highly dependent on the size of the jump of the optical electric field
intensity generating it, with six distinct DSW types identified \cite{saleh}.

As stated above, in the highly nonlocal limit the nematic DSW is of KdV-type with the DSW having
positive polarity. However, in the limit of weak nonlocality,
the nematic equations reduce to the NLS equation \cite{PR,conti2}, and
the nematic DSW becomes the NLS DSW, which is non-resonant.  The degree of nonlocality of the
optical response of a nematic is inversely proportional to the power of the optical beam, with the
response being highly nonlocal for lower power beams, transitioning to local as the beam power
increases \cite{PR,conti2}, as will be detailed in Section \ref{s:nemeqn}.  In this work, the
evolution of the nematic DSW structure as the degree of nonlocality ranges from highly nonlocal
(low-power beams) \cite{nembore,nemgennady,saleh} to local (high-power beams) will be studied.
As the nonlocality decreases, the changes in the DSW structure from those previously found
\cite{saleh} in the limit of high nonlocality to the standard NLS DSW \cite{gennady} will be
identified and the solutions for these will be derived. It is found that there
exist two additional DSW regimes over those for large nonlocality, including the NLS DSW
for zero nonlocality. The new DSW regime is a transition between the KdV DSW behavior for large
nonlocality and the NLS DSW behavior for very small nonlocality.  In this regime, the DSW structure
consists of a resonant wavetrain headed by a partial DSW which takes the solution to the initial
level ahead, similar to the resonant DSW for the KdV equation with fifth-order dispersion
\cite{pat,negkdv}.  As the nonlocality decreases, the optical power increases, the resonant
wavetrain contracts with the leading partial DSW expanding and becoming a full NLS DSW.  The
analytical solutions for the various DSW types will be compared with full numerical solutions
of the nematic equations.

\section{Nematic Equations}
\label{s:nemeqn}

\begin{figure}
\centering
\includegraphics[width=0.7\textwidth]{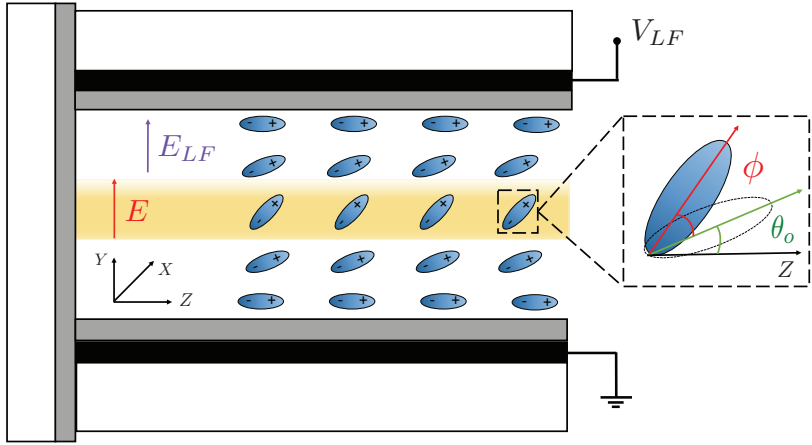}
\caption{Sketch of nematic cell.  A coherent light beam (yellow region) whose electric field $E$ is polarised in the $Y$ direction propagates in the $Z$ direction through
a cell filled with a dye-doped nematic liquid crystal.  Thin film electrodes (black) are deposited on the upper and lower cell walls (grey).  An external low frequency voltage bias $V_{LF}$ creates an electric field $E_{LF}$ to pre-tilt the molecules at an angle $\theta_{0}$ to $Z$. The nematic
molecules which are located at the boundaries are held tightly by the virtue of the
anchoring films. The far right inset (black) dashed box exhibits the angular rotation of a
nematic molecule with respect to the propagation direction  $Z$ in the absence ($\theta_{0}$) and presence ($\theta_{0} + \phi$) of the optical beam.}
\label{f:sketch}
\end{figure}

Let us consider the propagation of a linearly, extra-ordinarily polarized, coherent light beam of
wavenumber $k_{0}$, wavelength $\lambda_{0} = 2\pi /k_{0}$, through a planar cell filled with nematic liquid crystals. The optical beam is
assumed to propagate down the cell along in the $Z$ direction, with its electric field $E$
polarised in the $Y$ direction. The coordinate $X$ then completes the coordinate system. Nematic
liquid crystals are a uniaxial medium consisting of elongated molecules, with the long axis
termed the molecular director.  The refractive index of the medium is $n_{\parallel}$ for optical
beams polarized along the molecular director and $n_{\perp}$ for fields polarized orthogonal
to the director.  A fundamental property of nematic liquid crystals is that the
feature the so-called Fre\'edericksz threshold, whereby a minimum optical power is needed to rotate
the nematic molecules and thus change the refractive index of the medium \cite{khoo}.
However, high optical powers lead to heating of the nematic medium, which can cause the loss
of the nematic state if the temperature change is high enough \cite{khoo,assantokhoo}.
One method to overcome this is to pre-tilt the nematic molecules at an angle $\theta_{0}$
with respect to the $Z$-direction upon the application of an external static
electric field $E_{LF}$, so that milli-Watt power beams can rotate the nematic molecules
\cite{assantokhoo}.  Let us denote the optically induced rotation of the nematic by $\phi$,
so that in the presence of an optical beam the total angle of the nematic director to the $Z$ direction is $\theta = \theta_{0} + \phi$.  This configuration of the nematic cell is illustrated in Figure \ref{f:sketch}.
The dimensional equations governing
the propagation of the optical beam in the nematic cell are then of the following form:
\begin{equation}
 2ik_{0} n_{e} \frac{\partial E}{\partial Z} + \nabla^{2} E
 + k_{0}^{2} \left[ n_{\perp}^{2} \cos^{2} \theta + n_{\parallel}^{2} \sin^{2} \theta - n_{\perp}^{2} \cos^{2} \theta_{0}
 - n_{\parallel}^{2} \sin^{2} \theta_{0} \right] E = 0,
 \label{e:dime}
 \end{equation}
for the electric field of the beam and
\begin{equation}
 K\nabla^{2} \phi + \left[\frac{1}{4}\epsilon_{0} \Delta \epsilon |E|^{2}
+ \frac{1}{2}\Delta \epsilon_{LF} E_{LF}^2\right] \sin 2\left( \theta_{0} + \phi\right) = 0, \label{e:dimdir}
\end{equation}
for the nematic response \cite{PR,Wiley,conti2,physdreview}. Here, the extraordinary refractive index of the nematic is:
\begin{equation}
 n_{e}^{2} = \frac{n_{\perp}^{2}n_{\parallel}^{2}}{n_{\parallel}^{2}\cos^{2}\theta
 + n_{\perp}^{2}\sin^{2} \theta}.
 \label{e:ne}
\end{equation}
In the above equations, $\Delta \epsilon = n_{\parallel}^{2} - n_{\perp}^{2}$ is the optical
anisotropy, $\Delta \epsilon_{LF}$ is the low-frequency dielectric anisotropy and $\epsilon_{0}$ is the electrical permittivity of free space.
In addition, the constant $K$ is the elastic medium constant in the one constant approximation
for which the elastic constants of bend, twist and splay are taken equal \cite{khoo,PR}.

The nematic equations (\ref{e:dime}) and (\ref{e:dimdir}) are highly nonlinear and difficult to
analyse.  However, for milli-Watt power beams the optical induced rotation $\phi$ is small,
$|\phi| \ll \theta_{0}$, so that these equations can be expanded in Taylor series
around $|\phi|$. 
In addition, these equations can be put in dimensionless form
using typical scales $L_{Z}$ down the cell and $W$ transverse to the down cell direction, 
as well as a typical scale $A_{b}$ for the electric field of the optical beam, so that
\begin{equation}
 Z = L_{z} z, \quad X = Wx, \quad Y = Wy, \quad E = A_{b}u.
 \label{e:nondim}
\end{equation}
Here, $(x,y,z)$ is the non-dimensional coordinate system and $u$ is the non-dimensional electric field of the optical beam.
The electric field scale is obtained by assuming that the input optical beam is a Gaussian
beam of power $P_{b}$, amplitude $A_{b}$ and width $W_{b}$, so that
\begin{equation}
 A_{b}^{2} = \frac{2P_{b}}{\pi \Gamma W_{b}^{2}}, \quad \Gamma = \frac{1}{2} \epsilon_{0} cn_{e}, \quad
 n_{e}^{2} = \frac{n_{\parallel}^{2}n_{\perp}^{2}}
 {n_{\parallel}^{2}\cos^{2} \theta_{0} + n_{\perp}^{2}
 \sin^{2} \theta_{0}}.
 \label{e:enondim}
\end{equation}
Substituting these into the nematic equations (\ref{e:dime}) and (\ref{e:dimdir}), and
expanding in Taylor series for small $|\phi|$, we find \cite{physdreview,timlocal} that suitable scalings are
\begin{equation}
 L_{Z} = \frac{4n_{e}}{\Delta \epsilon k_{0}\sin 2\theta_{0}},
 \quad
 W = \frac{2}{k_{0} \sqrt{\Delta \epsilon \sin 2\theta_{0}}},
\label{e:azw}
\end{equation}
and the resulting non-dimensional equations read
\begin{eqnarray}
 i \frac{\partial u}{\partial z} + \frac{1}{2}\nabla^{2} u + 2\phi u & = & 0 ,
\label{e:eeqn} \\
\nu \nabla^{2} \phi - 2q\phi & = & - 2|u|^{2} . \label{e:direqn}
\end{eqnarray}
Here, the dimensionless elasticity and pre-tilting parameters, $\nu$ and $q$, are given by
\begin{equation}
\nu = \frac{8K}{\epsilon_{0} \Delta \epsilon A_{b}^{2} W^{2} \sin 2\theta_{0}} =
\frac{\pi K\Gamma k_{0}^{2}W_{b}^{2}}{\epsilon_{0} P_{b}}, \quad q = \frac{4\Delta \epsilon_{LF}E_{LF}^{2}
\cos 2\theta_{0}}{\epsilon_{0} \Delta \epsilon A_{b}^{2} \sin 2\theta_{0}}.
\label{e:nu}
\end{equation}
Typical experimental beam parameter values are power $P_{b} = 2mW$ and half-width
$W_{b} = 1.5 \mu m$, with a wavelength $\lambda_{0} = 2\pi/k_{0} = 1.064\mu m$ in the near
infrared \cite{PR,physdreview}.  For the liquid crystal E7, a typical
elastic constant is $K = 1.2\times 10^{-11} N$. These parameter values give the elasticity
parameter $\nu = O(100)$, as found in other studies \cite{physdreview,waveguide,yana}. This high value of $\nu$
means that the nematic is operating in the highly nonlocal regime, in that the elastic response of
the nematic extends far beyond the waist of the optical beam \cite{PR,Wiley, conti2}.
However, $\nu$ is inversely proportional to the beam power $P_{b}$.  Note that for $\nu=0$ the
nematic equations (\ref{e:eeqn}) and (\ref{e:direqn}) reduce to the standard NLS equation
\begin{equation}
 i\frac{\partial u}{\partial z} + \frac{1}{2} \nabla^{2} u + \frac{2}{q} |u|^{2} u = 0.
 \label{e:nls}
\end{equation}
This is the local response limit for the nematic.  Note that in $(2+1)$-dimensions,
beams governed by this equation are unstable and can show catastrophic collapse above
a critical power \cite{kivshar}. It is known that a nonlocal response, $\nu$ large, stabilizes
$(2+1)$-dimensional optical beams \cite{PR,Wiley,conti2,physdreview}.
This is because the nematic response equation~(\ref{e:dimdir}) is elliptic and so its solution
depends on $u$ in the entire domain, the origin of the physical concept of nonlocality. Hence, by
adjusting the beam power $P_{b}$, the response of the medium can be adjusted from nonlocal to
local, as long as the induced heating does not destroy the nematic phase at high power.

The nematic system (\ref{e:eeqn}) and (\ref{e:direqn}) is a focusing NLS-type system,
that is the refractive index in the dimensional equation (\ref{e:dime}) increases with beam
intensity $|u|^{2}$. Since focusing NLS equations do not possess (stable) DSW solutions, the
equation needs to be defocusing; in such a case, the refractive index decreases with beam
intensity, and DSW solutions do
exist \cite{elreview}.  The nematic medium can
feature a defocusing response through the addition of azo-dyes \cite{gaetanodark}.
The change in the nematic response due to the addition of the azo-dye is physically
complicated, with the ``order parameter'' change being opposite to that without the
presence of the dye.  A simple model of this response change is to
modify the electric field equation (\ref{e:eeqn}) from focusing to defocusing.  In addition,
the analysis of DSWs is simplest in $(1+1)$-dimensions as then there are no geometric spreading effects. With these assumptions and simplifications, the nematic equations become
\begin{eqnarray}
 i \frac{\partial u}{\partial z} + \frac{1}{2}\frac{\partial^{2} u}{\partial x^{2}}
 - 2\phi u & = & 0,
\label{e:eeqnd} \\
\nu \frac{\partial^{2}\phi}{\partial x^{2}} - 2q\phi & = & - 2|u|^{2}.
\label{e:direqnd}
\end{eqnarray}
The same system of equations also describes optical beam propagation in thermal optical media for which the refractive index depends on the temperature of the medium \cite{gho}.  Such thermal optical media typically have a defocusing response.
For these defocusing nematic equations, a suitable initial condition which will generate a DSW is the intensity jump
initial condition
\begin{equation}
 u = \left\{ \begin{array}{cc}
              u_{-}, & x < 0 \\
              u_{+}, & x > 0
             \end{array}
     \right. , \qquad
     \phi = \left\{ \begin{array}{cc}
              \frac{u_{-}^{2}}{q}, & x < 0 \\
              \frac{u_{+}^{2}}{q}, & x > 0
             \end{array}.
     \right.
\label{e:ic}
\end{equation}
The DSW solution of the defocusing nematic equations (\ref{e:eeqnd}) and (\ref{e:direqnd}) has been
studied in the highly nonlocal limit $\nu \gg 1$ in previous work \cite{nembore,nemgennady,saleh}.
As stated above, in the local limit $\nu \ll 1$ this system reduces to a perturbed defocusing NLS
equation and becomes the defocusing NLS equation for $\nu = 0$. The DSW solution of the
defocusing NLS equation is known \cite{gennady}, so that the perturbed local DSW solution can be
found using perturbed Whitham modulation theory \cite{kamchat}. Before studying the behaviour of
the nematic DSW as $\nu$ varies from large to small, some previously derived results
\cite{nembore,nemgennady,saleh} will be briefly summarized.

The analytical DSW solutions derived in this Section will be compared with full numerical solutions
of the nematic equations (\ref{e:eeqnd}) and (\ref{e:direqnd}) with the initial condition
(\ref{e:ic}).  The electric field equation (\ref{e:eeqnd}) will solved using the pseudo-spectral
method of Fornberg and Whitham \cite{bengt}, as extended \cite{chan,tref} to improve the stability
for high wavenumbers through the use of an integrating factor.  The $x$ derivatives are calculated
using the Fast Fourier Transform (FFT) and the solution is advanced in $z$ in Fourier space
employing the fourth-order Runge-Kutta method, as detailed in previous work \cite{saleh,salehthesis}.
The step initial condition (\ref{e:ic}) was smoothed using the hyperbolic tangent function,
as detailed in \cite{saleh,salehthesis}. The director equation (\ref{e:direqnd}) was also solved using the
FFT, with does not have a singularity at zero wavenumber due to the $2q\theta$ term.

\begin{figure}
\centering
 \includegraphics[width=0.25\textwidth,angle=270]{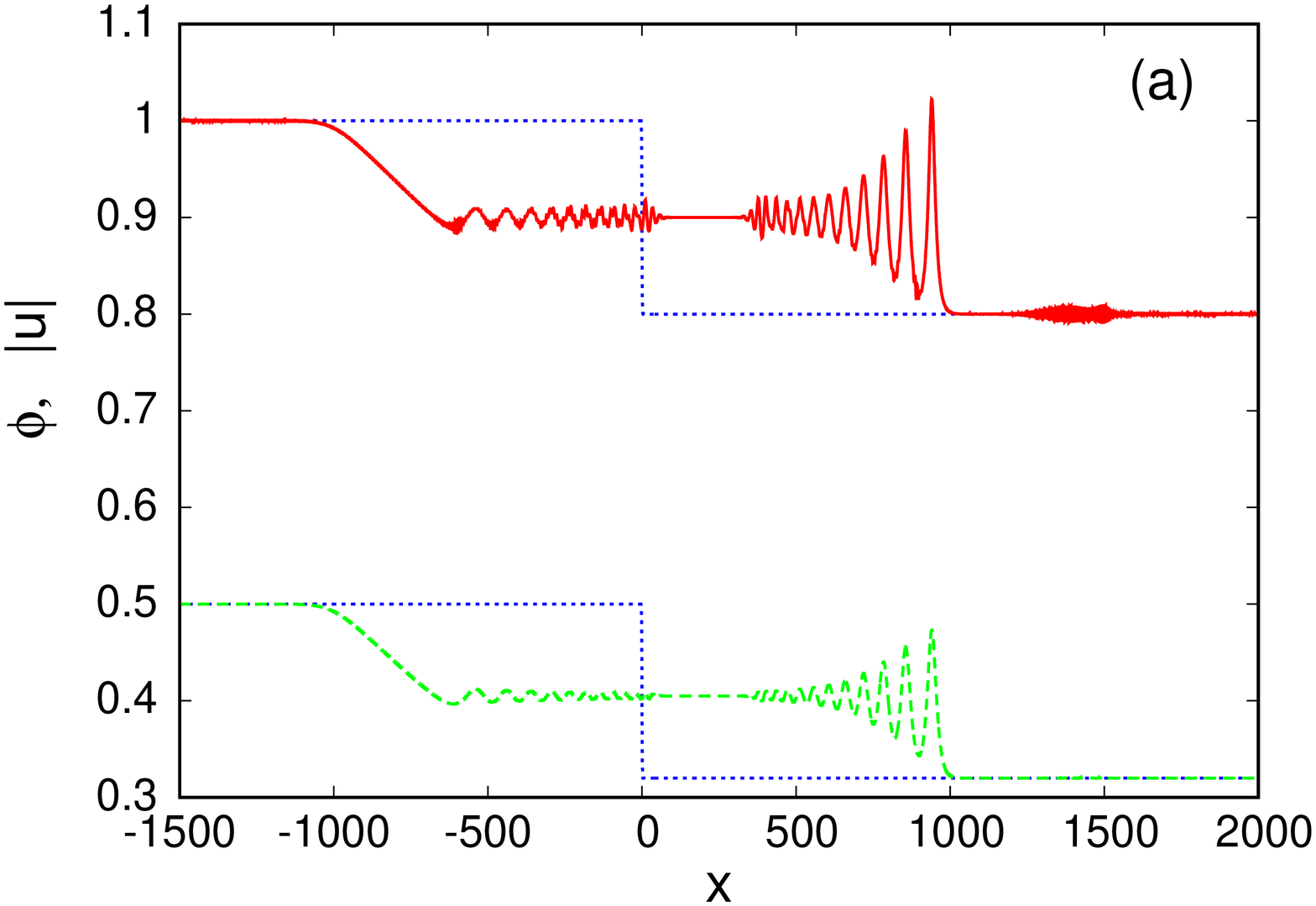}
 \includegraphics[width=0.25\textwidth,angle=270]{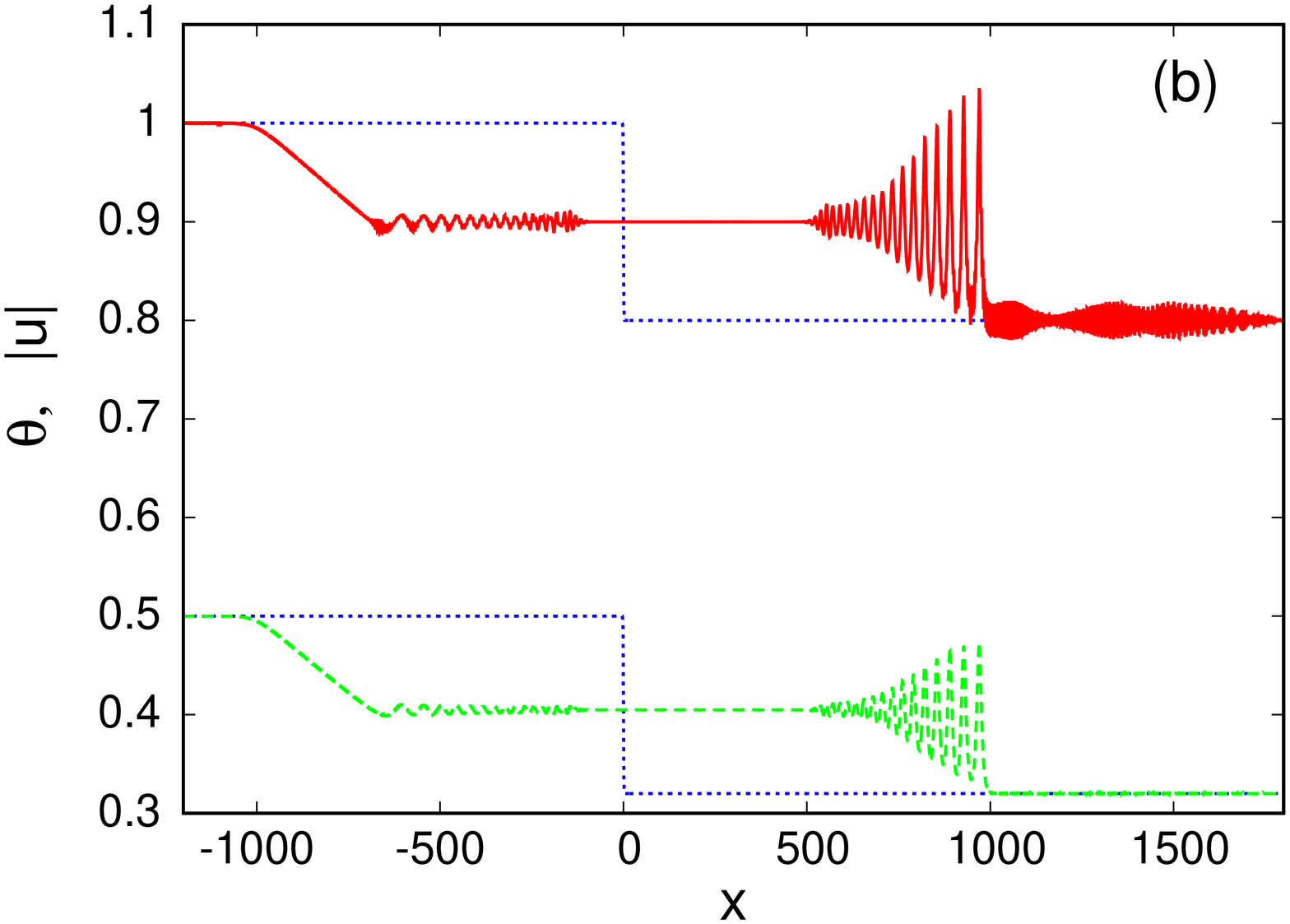}
 \includegraphics[width=0.25\textwidth,angle=270]{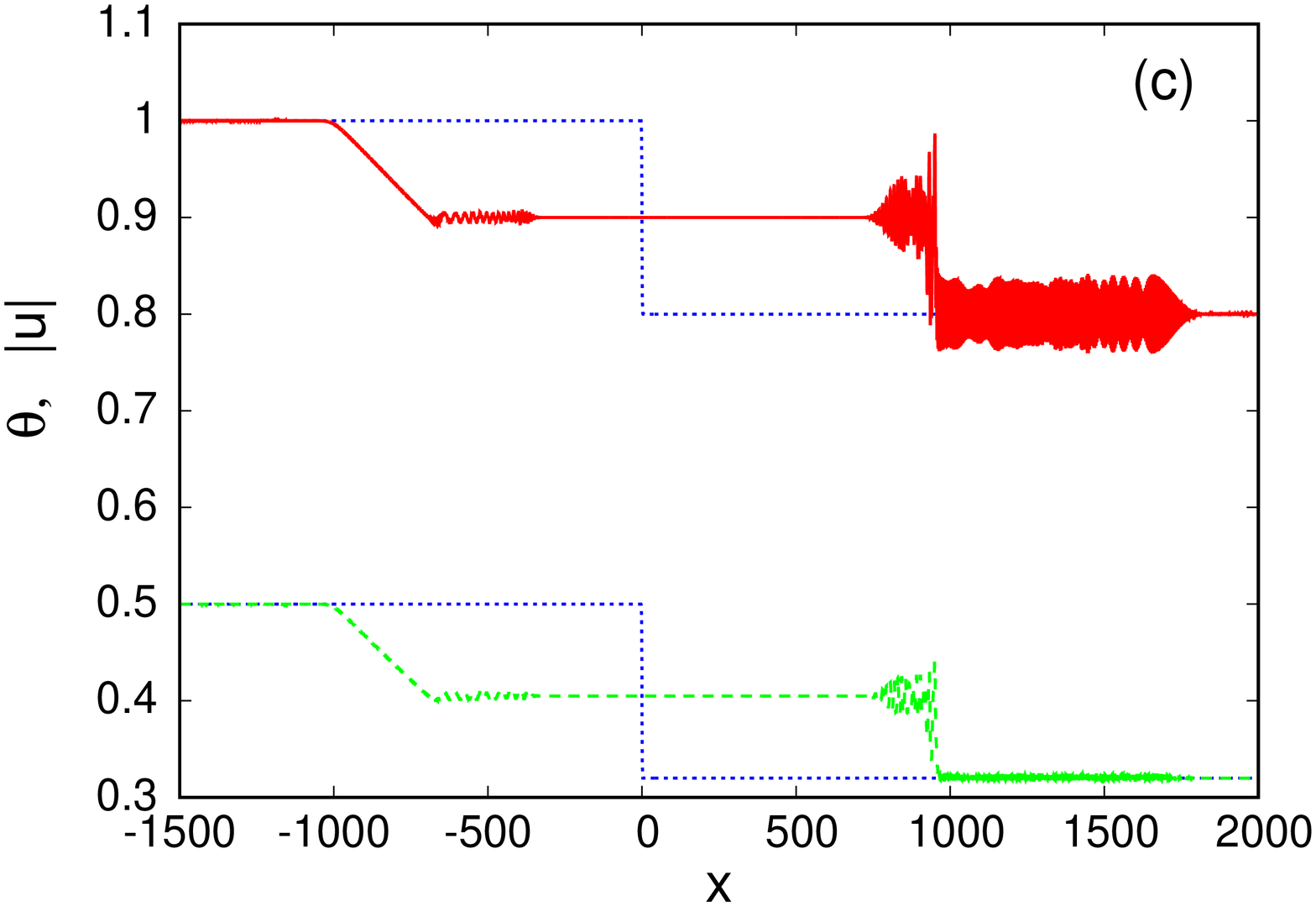}
 \includegraphics[width=0.25\textwidth,angle=270]{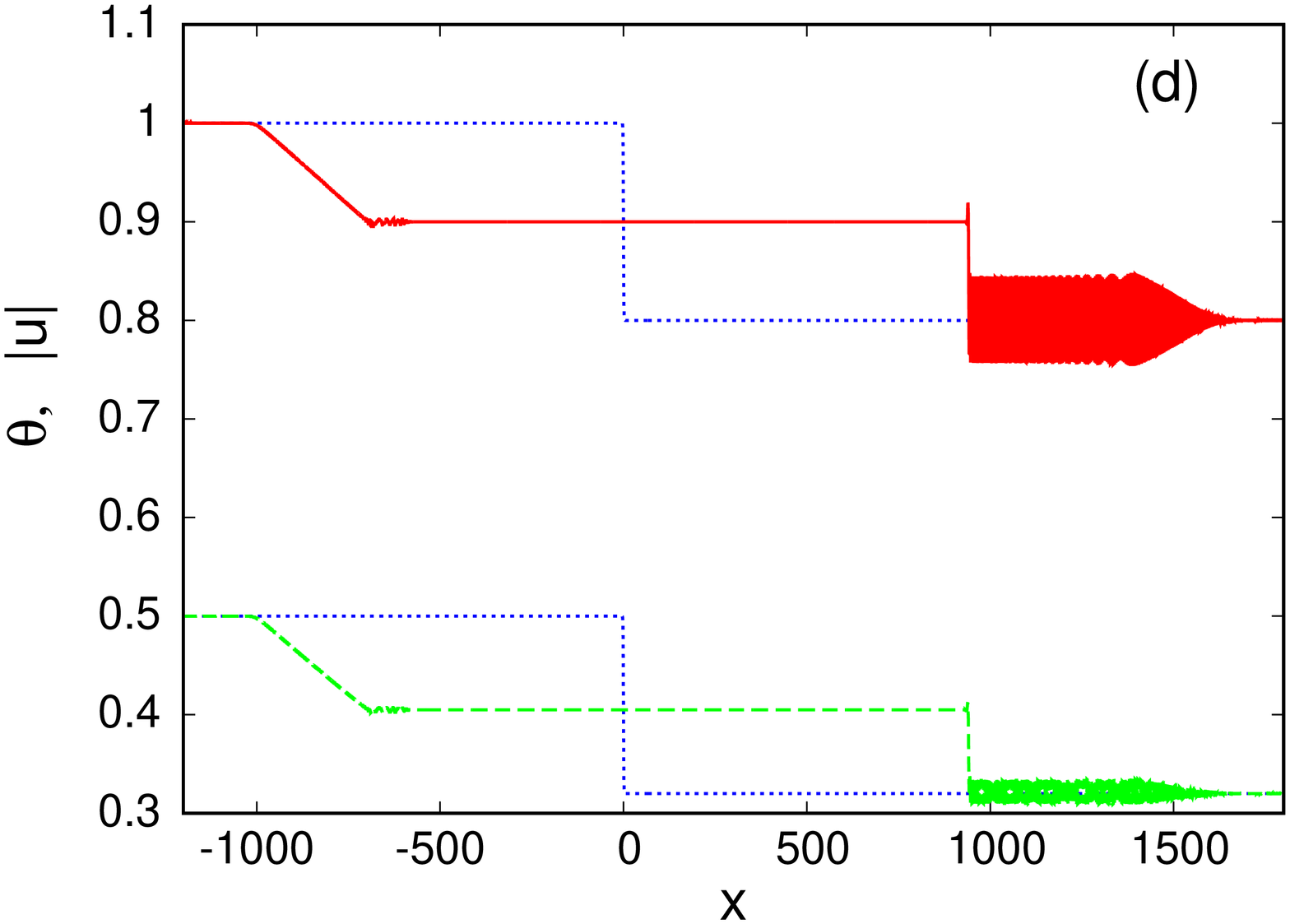}
 \includegraphics[width=0.25\textwidth,angle=270]{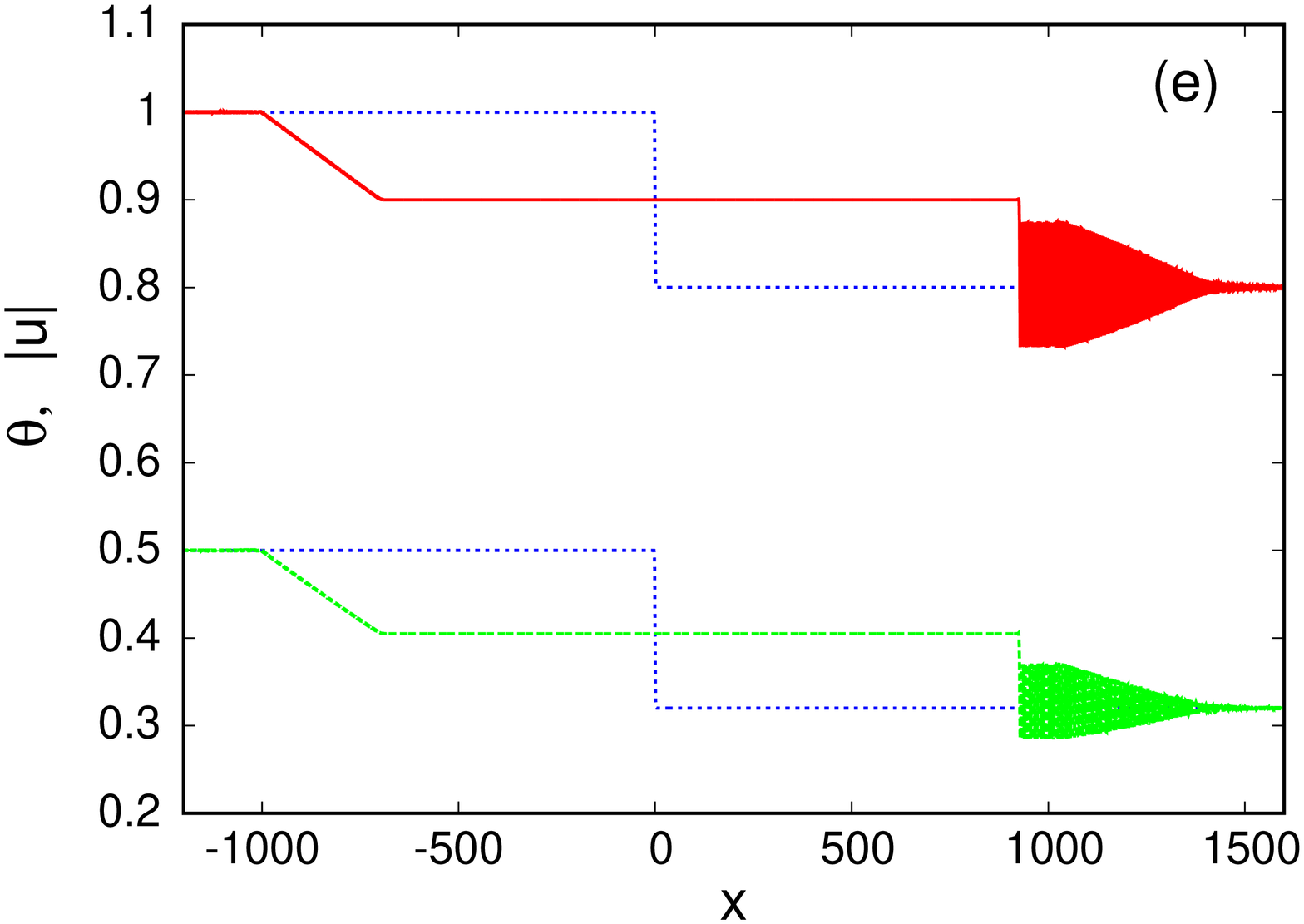}
 \includegraphics[width=0.25\textwidth,angle=270]{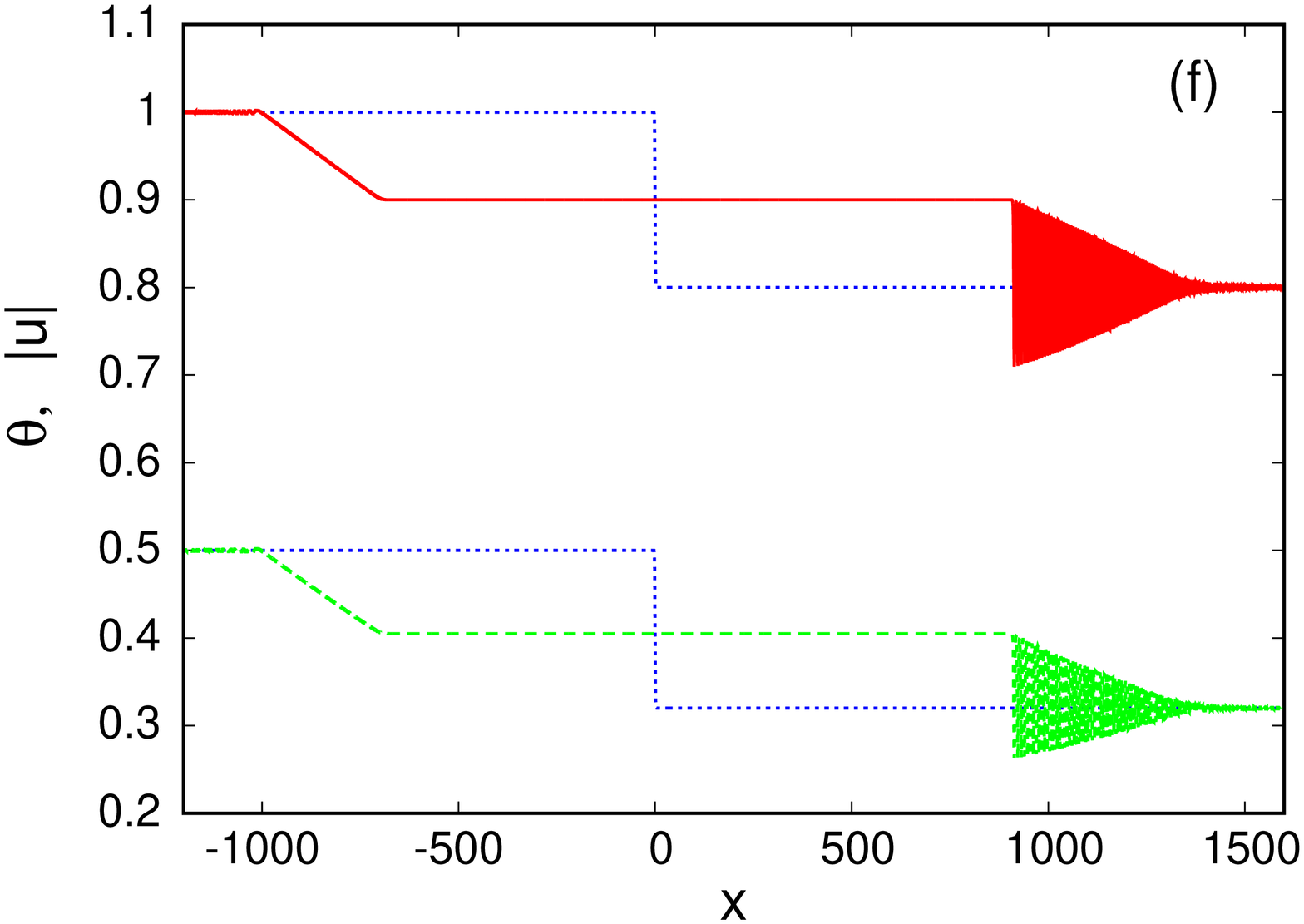}
\caption{Numerical solutions of nematic equations (\ref{e:eeqnd}) and (\ref{e:direqnd}) for
initial condition (\ref{e:ic}) with $u_{+} = 0.8$ and $u_{-} = 1.0$.  Red (solid) lines:  $|u|$ at
$z=1000$; green (dashed) lines: $\phi$ at $z=1000$; blue (dotted) lines:  $|u|$ at $z=0$ (upper)
and $\phi$ at $z=0$ (lower). (a) PDSW with $\nu = 200$, (b) RDSW with $\nu = 40$,
(c) CDSW with $\nu = 10$, (d) TDSW with $\nu = 3.0$,
(e) resonant NLS type DSW with $\nu = 1.0$, (f) NLS type DSW with $\nu = 0.5$.  Here $q =2$.}
\label{f:types}
\end{figure}

Figure~\ref{f:types} displays a summary of the nematic DSW types as the nonlocality $\nu$ varies
from large $\nu$, corresponding to a highly nonlocal response for low beam power, to small $\nu$,
pertinent to a local medium response for high beam power.  The terminology for the DSW regimes
will be taken from previous work on radiating DSWs \cite{patkdv,saleh}. In particular, the DSW
regimes from nonlocal to local response will now be detailed, for $u_{-} = 1$ and $u_{+} = 0.8$,
with the existence intervals for the various DSW types given in Table~\ref{t:typebounds}.

\begin{itemize}

 \item  PDSW (perturbed DSW):  This regime is illustrated in Figure \ref{f:types}(a).  The nematic DSW is essentially a KdV DSW governed by the Kawahara equation (\ref{e:kdv5nem}) and its solution can be found as a perturbed KdV DSW \cite{saleh,perturbkdv}.  This DSW regime is non-resonant and there is no resonant wavetrain attached to the leading edge of the DSW.

 \item  RDSW (radiating DSW):  As the nonlocality $\nu$ decreases, the DSW becomes resonant with a resonant wavetrain attached to the leading edge of the DSW, as illustrated in Figure~\ref{f:types}(b).  As all individual waves in the DSW are resonant, resonant waves are emitted from the DSW, which results in the DSW not being rank ordered \cite{saleh}.

 \item  CDSW (crossover DSW):  As the nonlocality decreases and the amplitude of the resonant wavetrain grows, the DSW becomes unstable with a total loss of the rank ordering of the waves of the DSW \cite{patkdv,saleh}.  This regime is illustrated in Figure~\ref{f:types}(c).

 \item  TDSW (travelling DSW):  As the amplitude of the resonant wavetrain grows, the shedding of conserved quantities into resonant radiation eventually destroys the DSW, leaving a high amplitude resonant wavetrain with a negative polarity solitary wave linking this wavetrain to the intermediate level \cite{patkdv} as seen in Figure~\ref{f:types}(d).
While there is a solitary wave linking the resonant wavetrain to the intermediate level, this linking can be conveniently
treated as a Whitham shock \cite{patjump}, a shock wave in the Whitham modulation equation
variables.  The resonant wavetrain is brought down to the level $u_{+}$ ahead by a partial DSW
\cite{saleh}.  This partial DSW has linear dispersive waves at its leading edge, but has a finite
wavelength wave at its trailing edge, that is, it is not bounded by solitary waves
 at the trailing edge \cite{negkdv,resflowmod}, as for a standard DSW.

\item  RNLS DSW (radiating NLS DSW):  Further decrease in the nonlocality
results in the amplitude of the linking solitary wave becoming negligible, so that the wave form
consists of a (stable) resonant wavetrain headed by a partial DSW which brings the wavetrain down
to the level $u_{+}$ ahead, as seen in Figure~\ref{f:types}(e). 
This DSW
regime does not occur in the high nonlocality limit as it is a ``bridge'' to the local NLS DSW
for $\nu = 0$.

 \item  NLS DSW:  As the nonlocality $\nu$ decreases
to $\nu = 0$ and the nematic equations (\ref{e:eeqnd}) and (\ref{e:direqnd}) reduce to the NLS
equation, the resonant wavetrain length contracts and the leading partial DSW evolves to a full DSW,
with linear dispersive waves at the leading edge and solitary waves at the trailing edge.
The resonant wavetrain then disappears and the leading DSW bringing the wavetrain down to the level $u_{+}$ ahead attaches to the intermediate level with $|u| = u_{i}$, as in Figure \ref{f:types}(f).  The resulting DSW is essentially an NLS DSW, which completes the transition from the KdV-type DSW for high nonlocality, that is for low power beams, to a NLS DSW for low nonlocality, that is for high power beams.

\end{itemize}

In addition to these six DSW types, when $u_{-} - u_{+}$ is large enough as $u_{+} \to 0$ there is an additional DSW type which is a sub-case of the TDSW regime, the vacuum DSW (VDSW) \cite{gennady}, for which the electric field $u$ of the resonant radiation vanishes at a point \cite{nemgennady,saleh}.  As the solution for this DSW type has been previously analysed \cite{saleh} and it is a sub-case of the TDSW regime, it will not be considered here.  In addition, the analytical work of this manuscript is based on $u_{-} - u_{+}$ being small, which is not valid in the VDSW
regime.

Table~\ref{t:typebounds} shows that the transition regimes of TDSW and RNLS DSW
which evolve the DSW from the KdV-type DSW for high nonlocality to the NLS-type DSW for weak
nonlocality exist for very restricted ranges of $\nu$, so that this transition is rapid.  Over most
of the range of $\nu$ the DSW is of KdV-type, one of the PDSW, RDSW and CDSW types.  As noted, the
first four DSW regimes also occur for the Kahawara equation and the nematic equations in the lower
power, high nonlocality limit.

\begin{table}
\centering
 \begin{tabular}{|c|c|} \hline
   PDSW       &  $88 < \nu$        \\ \hline
   RDSW       &  $34 < \nu < 88$   \\ \hline
   CDSW       &  $4.1 < \nu < 34$  \\ \hline
   TDSW       &  $1.53 < \nu < 4.1$ \\ \hline
   RNLS DSW    &  $0.60 < \nu < 1.53$ \\ \hline
   NLS DSW     & $0 \le \nu < 0.60$ \\ \hline
 \end{tabular}
 \caption{Regime boundaries for type classifications of Figure \ref{f:types}.
 Here $u_{-} = 1.0$, $u_{+} = 0.8$ and $q=2$.}
 \label{t:typebounds}
\end{table}

The standard method to analyse DSW solutions of nonlinear dispersive wave equations of
the defocusing NLS-type is to transform the equation into the hydrodynamic
form using the Madelung transformation \cite{elreview}
\begin{equation}
 u = \sqrt{\rho} e^{i\psi}, \quad v = \psi_{x},
\label{e:med}
\end{equation}
where the real functions $\rho$ and $\psi$ denote the density and phase of the
field $u$, while $v$ is the fluid velocity. Upon substituting, it is found
that the nematic equations (\ref{e:eeqnd}) and (\ref{e:direqnd}) become
\begin{eqnarray}
 \frac{\partial \rho}{\partial z} + \frac{\partial }{\partial x} \left( \rho v \right) & = & 0, \label{e:mass} \\
 \frac{\partial v}{\partial z} + v \frac{\partial v}{\partial x} + 2\frac{\partial \phi}{\partial x}
- \frac{\partial}{\partial x} \left( \frac{\rho_{xx}}{4\rho} - \frac{\rho_{x}^{2}}{8\rho^{2}} \right) & = & 0,
\label{e:mom} \\
\nu \frac{\partial^{2}\phi}{\partial x^{2}} - 2q\phi & = & -2\rho. \label{e:thm}
\end{eqnarray}
The nematic equations are characterized by
the linear dispersion relation \cite{nembore,nemgennady}
\begin{equation}
 \omega = k\bar{v} + \frac{\sqrt{\bar{\rho}k}}{\sqrt{\nu k^{2} + 2q}} \left[
\frac{\nu k^{2} + 2q}{4\bar{\rho}} k^{3} + 4k \right]^{1/2},
\label{e:disp}
\end{equation}
for waves around the mean level $\bar{\rho}$ for $\rho$ and $\bar{v}$ for $v$.
In the short wave and high nonlocality limit, $\nu k^{2} \gg 1$, this dispersion relation can be approximated by
\begin{equation}
 \omega = k\bar{v} + \frac{1}{2} k^{2} + \cdots.
\label{e:displargenu}
\end{equation}
In the opposite local limit with $\nu$ small, the dispersion relation can be expanded as
\begin{equation}
 \omega = k\bar{v} + k^{2} \left[\frac{1}{4}k^{2} + \frac{2}{q} \bar{\rho}\right]^{1/2}
 - \frac{\nu \bar{\rho}k^{3}}{2q^{2}} \left[ \frac{1}{4}k^{2}
 + \frac{2}{q} \bar{\rho} \right]^{-1/2} + \cdots.
 \label{e:dispnusmall}
\end{equation}
As expected, at the leading-order, $O(1)$, this dispersion relation is the same
as that for the NLS equation \cite{elreview,chaos}. These dispersion relations are
needed for the determination of the resonant wavetrain generated by the nematic DSW by which the
linear phase velocity is matched to the velocity of the DSW front.

As seen from Figure~\ref{f:types}, the solution outside of the DSW and the resonant wavetrain is non-dispersive. On neglecting dispersion, the nematic equations (\ref{e:mass})--(\ref{e:thm}) become the shallow water equations \cite{whitham} with $\rho$ playing the role of fluid depth and $v$ the (horizontal) fluid velocity. In Riemann invariant form, the dispersionless nematic
equations read:
\begin{eqnarray}
 v + \frac{2\sqrt{2}}{\sqrt{q}} \sqrt{\rho} = R_{+} = \mbox{constant} & \mbox{on} & C_{+}:
\frac{dx}{dz} = V_{+} = v + \frac{\sqrt{2}}{\sqrt{q}} \sqrt{\rho}
\label{e:cp} \\
v - \frac{2\sqrt{2}}{\sqrt{q}} \sqrt{\rho}= R_{-} = \mbox{constant} & \mbox{on} & C_{-}:
\frac{dx}{dz} = V_{-} = v - \frac{\sqrt{2}}{\sqrt{q}} \sqrt{\rho} ,
\label{e:cm}
\end{eqnarray}
The initial level behind $u_{-}$ is linked to the intermediate shelf by a simple wave on the characteristic $C_{-}$.  This simple wave solution has been derived previously \cite{nembore,nemgennady,saleh} and is of the form:
\begin{equation}
|u| = \sqrt{\rho} = \left\{ \begin{array}{cc}
                        u_{-}, & \frac{x}{z} < -\frac{\sqrt{2}u_{-}}{\sqrt{q}} \\
                        \frac{\sqrt{q}}{3\sqrt{2}} \left[ \frac{2\sqrt{2}u_{-}}{\sqrt{q}}
                        - \frac{x}{z} \right], &
-\frac{\sqrt{2}u_{-}}{\sqrt{q}} \le \frac{x}{z} \le \frac{\sqrt{2}}{\sqrt{q}} \left( 2u_{-}
- 3\sqrt{\rho_{i}}
\right), \\
\sqrt{\rho_{i}}, & \frac{\sqrt{2}}{\sqrt{q}} \left( 2u_{-} - 3\sqrt{\rho_{i}} \right) < \frac{x}{z} \le s_{i}
\end{array}
\right.
\label{e:midsolnu}
\end{equation}
with $v = 2\sqrt{2}(u_{-} - \sqrt{\rho})/\sqrt{q}$, where $s_{i}$ is the velocity of the trailing
edge of the DSW which lies on the intermediate level $u_{i}$. This level can be determined by the
requirement that the Riemann invariant along the characteristics $C_{-}$, that is, $R_{-}$, is
conserved across the nematic DSW \cite{nembore}, giving
\begin{equation}
 u_{i}=\frac{1}{2}\left(u_{-}+u_{+}\right).
\label{e:shelf}
\end{equation}
We can see from the above calculations that the phase gradient on the intermediate
level $v_{i}$ is then
\begin{equation}
    v_{i}=\frac{2\sqrt{2}}{\sqrt{q}}(u_{-}-\sqrt{\rho_{i}}).
    \label{e:v2expr}
\end{equation}

In the small jump limit $|u_{-} - u_{+}| \ll 1$ the nematic equations (\ref{e:eeqnd}) and
(\ref{e:direqnd}) can be reduced --in the high nonlocality regime ($\nu\gg 1$) under
consideration-- to a KdV equation with fifth-order dispersion \cite{nemgennady,horikis}.
This will be justified below upon employing a multiscale expansion method.


\subsection{Derivation of the extended KdV and Kawahara equations}

We seek solutions of Eqs.~(\ref{e:med})--(\ref{e:mom}) in the form of the following asymptotic
expansions in the formal small parameter $\varepsilon \equiv \sqrt{u_{i} - u_{+}}$ (with
$0<\varepsilon\ll 1$)
\begin{align}
 |u|^2 &= \rho =  \rho_{+} + \varepsilon^{2} \rho_{1}(\xi,\eta) + \varepsilon^{4}\rho_{2}(\xi,\eta)
 + \cdots,
 \label{e:ukdv} \\
 v &= {\varepsilon ^2}{V_1} + {\varepsilon ^4}{V_2} + {\varepsilon ^6}{V_3} +  \cdots
\label{e:phikdv} \\
 \phi &= \frac{{\rho_{+}}}{q} + {\varepsilon ^2}{\phi _1} + {\varepsilon ^4}{\phi _2} + {\varepsilon ^6}{\phi _3} +  \cdots,
\label{e:thetakdv}
\end{align}
where $\rho_{+}=u^{2}_{+}$, and the unknown functions $\rho_j$, $V_j$ and $\phi_j$ ($j=1,2,3,\ldots$) depend on the stretched variables
\begin{equation}
\xi=\varepsilon (x-Uz), \quad \eta=\varepsilon^{3}z.
\label{scales}
\end{equation}
Here, $U$ will be treated as an unknown velocity, which will be determined self-consistently.

Substituting the expansions~(\ref{e:ukdv})--(\ref{e:thetakdv}) into Eqs.~(\ref{e:med})--(\ref{e:mom}),
and using the stretched coordinates~(\ref{scales}), we obtain a set of equations
at the different orders in $\varepsilon$. In particular, at the leading order, we derive
the following linear equations:
\begin{equation}
O({\varepsilon ^2}):\quad {\rho_1} - q{\phi _1} = 0
\end{equation}
and
\begin{equation}
  {O({\varepsilon ^3}):} \qquad
  U{V_{1\xi }} - 2{\phi _{1\xi }} = 0 \quad \mbox{and} \quad
  U{\rho_{1\xi }} - u_ + ^2{V_{1\xi }} = 0,
\label{e:v1}
\end{equation}
where subscripts denote partial derivatives. The compatibility of the above equations
suggests that the squared velocity $U^2$ is given by $U^2=2u_+^2/q$.
Next, Eqs.~(\ref{e:med})--(\ref{e:mom}) yield a set of nonlinear equations, namely:
\begin{equation}
O({\varepsilon ^4}):\quad 2{\rho_2} - 2q{\phi _2} + \nu {\phi _{1\xi \xi }} = 0
\label{l1}
\end{equation}
and
\begin{equation}
\begin{array}{cc}
  {O({\varepsilon ^5}):}&\begin{gathered}
   - u_ + ^2{V_{1\eta }} + 3U{\rho_1}{V_{1\xi} } - u_ + ^2{V_1}{V_{1\xi }} + Uu_ + ^2{V_{2\xi }} - 6{\rho_1}{\phi _{1\xi }} - 2u_ + ^2{\phi _{2\xi }} + \frac{1}{4}{\rho_{1\xi \xi \xi }} = 0 \\
  {\rho_{1\eta }} + {({\rho_1}{V_1})_\xi } - U{\rho_{2\xi }} + u_ + ^2{V_{2\xi }} = 0
\end{gathered}
\end{array}
\label{l2}
\end{equation}
The compatibility condition at this order can be found upon eliminating the fields $\rho_2$, $V_2$
and $\phi_2$ upon using Eqs.~(\ref{l1})--(\ref{l2}) and the definition of the velocity $U$. This
yields the following KdV equation:
\begin{equation}
{\rho_{1\eta }} + \frac{3}{2u_+}\sqrt {\frac{2}{q}} {\rho_1}{\rho_{1\xi }} + {\left( {\frac{1}{q}} \right)^{3/2}}\frac{{4\nu u_ + ^2 - {q^2}}}{{8\sqrt 2 {u_ + }}}{\rho_{1\xi \xi \xi }} = 0.
\label{kdv}
\end{equation}
To the next order of approximation, we obtain:
\begin{equation}
O({\varepsilon ^6}):\quad 2{\rho_3} - 2q{\phi _3} + \nu {\phi _{2\xi \xi }} = 0
\label{ho1}
\end{equation}
and
\begin{eqnarray}
  {O({\varepsilon ^7}):} & & 
\mbox{}   - 3u_ + ^4{\rho_1}{V_{1\eta }} - u_ + ^6{V_{2\eta }} + 3Uu_ + ^2(\rho_1^2 + u_ + ^2{\rho_2}){V_{1\xi }} - 3u_ + ^4{\rho_1}{V_1}{V_{1\xi }} \nonumber \\
  & & \mbox{} - u_ + ^6{V_2}{V_{1\xi }} + 3Uu_ + ^4{\rho_1}{V_{2\xi }} - u_ + ^6{V_1}{V_{2\xi }} + Uu_ + ^6{V_{3\xi }} - 6u_ + ^2\rho_1^2{\phi _{1\xi }} \nonumber \\
  & & \mbox{} - 6u_ + ^4{\rho_2}{\phi _{1\xi }} - 6u_ + ^4{\rho_1}{\phi _{2\xi }} - 2u_ + ^6{\phi _{3\xi }} - \frac{1}{2}u_ + ^2{\rho_{1\xi }}{\rho_{1\xi \xi }} + \frac{1}{2}u_ + ^2{\rho_1}{\rho_{1\xi \xi \xi }} \nonumber \\
  & & \mbox{} + \frac{1}{4}u_+ ^4{\rho_{2\xi \xi \xi }} = 0
  \label{ho1a_new}
  \quad \mbox{and} \\
  & & {\rho_{2\eta }} + {({\rho_1}{V_2} + {\rho_2}{V_1})_\xi } - U{\rho_{3\xi }} + u_ + ^2{V_{3\xi }} = 0.
\label{ho2}
\end{eqnarray}
It is now possible to follow the procedure used at the previous
order and eliminate the fields $\rho_3$, $V_3$
and $\phi_3$ from Eqs.~(\ref{ho1})--(\ref{ho2}). Indeed, solving Eq.\ (\ref{ho1}) for $\phi_3$, Eq.\ (\ref{ho2}) for $\rho_{3\xi}$ and
substituting into Eq.\ (\ref{ho1a_new}) eliminates every term with index
3 (recall $U^2=2u_+^2/q$). Furthermore, employing the equations obtained at the previous orders, we can express the fields $\phi_{1,2}$ and $V_{1,2}$ in
terms of the amplitudes $\rho_1$ and $\rho_2$, which yields
\begin{gather}
\frac{1}{{2U}}\int {{\rho _{1\eta\eta}}} \;d\xi + {\rho _{1\xi}}\int {{\rho _{1\eta}}} \;d\xi + \frac{5}{2}{\rho _1}{\rho _{1\eta}} + {\rho _{2\eta}} + 3U\rho _1^2{\rho _{1\xi}} + \frac{{3c}}{2}{({\rho _1}{\rho _2})_\xi} + \frac{{U\nu }}{{2q}}{\rho _{1\xi}}{\rho _{1\xi\xi}}\nonumber\\
 - \frac{{q - 2{U^2}\nu }}{{8{U^2}q}}{\rho _{1\xi\xi\eta}} - \frac{{q - 3{U^2}\nu }}{{4Uq}}{\rho _1}{\rho _{1\xi\xi\xi}} - \frac{{q - 2{U^2}\nu }}{{8Uq}}{\rho _{1\xi\xi\xi}} + \frac{{U{\nu ^2}}}{{8{q^2}}}{\rho _{1\xi\xi\xi\xi\xi}} = 0.
\label{hor_new}
\end{gather}
To this end, we multiply Eq.~(\ref{hor_new}) by $\varepsilon^2$ and add it to the KdV equation Eq.~(\ref{kdv}). Then, introducing the combined amplitude function
\begin{equation}
P=\rho_1+\varepsilon^2 \rho_2,
\end{equation}
we solve for $\rho_1=P-\varepsilon \rho_2$ and substitute the result into the above equation (\ref{hor_new}).
We hence obtain the nonlinear evolution equation for the field $P(\xi,\eta)$
\begin{gather}
  {P_\eta } + \frac{3}{2u_+}\sqrt {\frac{2}{q}} P{P_\xi } + {\left( {\frac{1}{q}} \right)^{3/2}}\frac{{4\nu u_ + ^2 - {q^2}}}{{8\sqrt 2 {u_ + }}}{P_{\xi \xi \xi }} \nonumber \\
\mbox{} + {\varepsilon ^2}\left( {{b_1}{P^2}{P_\xi } + {b_2}{P_\xi }{P_{\xi \xi }} + {b_3}P{P_{\xi \xi \xi }} + {b_4}{P_{\xi \xi \xi \xi \xi }}} \right) = 0.
  \label{Pofq_P}
\end{gather}
The coefficients $b_j$ ($j=1,2,3,4$) appearing in Eq.~(\ref{Pofq}) are given by
\begin{align*}
 {b_1} &=  - \frac{3}{{8u_ + ^3}}\sqrt {\frac{2}{q}},\quad
  {b_2} =  - {\left( {\frac{1}{q}} \right)^{3/2}}\frac{{20\nu u_ + ^2 - 13{q^2}}}{{32\sqrt 2 u_ + ^3}}, \\
  {b_3} &= {\left( {\frac{1}{q}} \right)^{3/2}}\frac{{4\nu u_ + ^2 + {q^2}}}{{16\sqrt 2 u_ + ^3}},\quad
  {b_4} = {\left( {\frac{1}{q}} \right)^{5/2}}\frac{{48{\nu ^2}u_ + ^4 + 8\nu {q^2}u_ + ^2 - {q^4}}}{{256\sqrt 2 u_ + ^3}}.
\end{align*}
Then, we seek an asymptotic expansion in the optical beam intensity $|u|$ as
\begin{equation}
    |u|=u_{+}+\varepsilon^{2} Q + \cdots,
\end{equation}
and use the relation $|u|=\sqrt \rho$. This asymptotically gives $P=2u_{+}Q$. The reductive nonlinear equation (\ref{Pofq_P}) can now be written in terms of the field $Q(\xi,\eta)$ as 
\begin{gather}
  {Q_\eta } + 3\sqrt {\frac{2}{q}} Q{Q_\xi } + {\left( {\frac{1}{q}} \right)^{3/2}}\frac{{4\nu u_ + ^2 - {q^2}}}{{8\sqrt 2 {u_ + }}}{Q_{\xi \xi \xi }} \nonumber \\
\mbox{} + {\varepsilon ^2}\left( {{c_1}{Q^2}{Q_\xi } + {c_2}{Q_\xi }{Q_{\xi \xi }} + {c_3}Q{Q_{\xi \xi \xi }} + {c_4}{Q_{\xi \xi \xi \xi \xi }}} \right) = 0.
  \label{Pofq}
\end{gather}
The coefficients $c_j$ ($j=1,2,3,4$) appearing in Eq.~(\ref{Pofq}) are given by
\begin{align*}
 {c_1} =  4u^{2}_{+}b_{1},\quad
  {c_2} = 2u_{+}b_{2}, \quad 
  {c_3} = 2u_{+}b_{3}, \quad c_{4}= b_{4}.
\end{align*}
Notice that Eq.~(\ref{Pofq}) is the so-called
extended KdV equation (eKdV), which can model the evolution of steeper waves, with shorter
wavelengths, than those governed by the KdV equation. As such, the eKdV equation has been used to describe solitary waves
in plasmas \cite{kodama} and shallow water waves \cite{TimNoel} in the presence of higher order
effects.  We note that the coefficient of the third derivative dispersive term changes sign when
\begin{equation}
 \nu = \frac{q^{2}}{4u_{+}^{2}}.
 \label{e:nlsbound}
\end{equation}
Hence, in the high nonlocality, low power, limit, such that $\nu > q^{2}/(4u_{+}^{2})$, the coefficient
of the third derivative in eKdV equation is positive, so that its DSW
(and solitary wave) solutions have positive polarity, with solitary waves at its leading edge
and linear dispersive waves at its trailing edge. On the other hand, in the local limit,
$\nu < q^{2}/(4u_{+}^{2})$, the coefficient of the third derivative is negative and the DSW has negative polarity, with linear dispersive waves at its leading edge and solitary waves at its trailing edge, so that it resembles the standard NLS DSW \cite{gennady}.  The nematic DSW then undergoes
a change of form from a KdV-type DSW to an NLS-type DSW as $\nu$ decreases at the value of the nonlocality parameter
given by (\ref{e:nlsbound}).

%
%
%

At this point, it is useful to make the following remarks.
First, in the highly nonlocal limit, the dominant higher-order coefficient is the one of
the fifth-order dispersion term, namely $c_4 \propto \nu^2$. Thus, in this limit, the
eKdV equation (\ref{Pofq_P}) may be approximated by the Kawahara equation
\begin{equation}
 {P_\eta } + \frac{3}{2u_+}\sqrt {\frac{2}{q}} P{P_\xi } + {\left( {\frac{1}{q}} \right)^{3/2}}\frac{{\left( {4u_+^2\nu  - {q^2}} \right)}}{{8\sqrt 2 {u_ + }}}{P_{\xi \xi \xi }}
  + {\varepsilon ^2} \frac{3}{16\sqrt{2}}\left( {\frac{1}{q}} \right)^{5/2}
  u_+ \nu^2 {P_{\xi \xi \xi \xi \xi }} = 0,
\label{e:kdv5nem}
\end{equation}
as was done in previous work \cite{nemgennady,saleh}.

\section{Nonlocal to local nematic DSWs}
\label{s:nonlocallocal}

\begin{figure}
\centering
 \includegraphics[width=0.5\textwidth,angle=270]{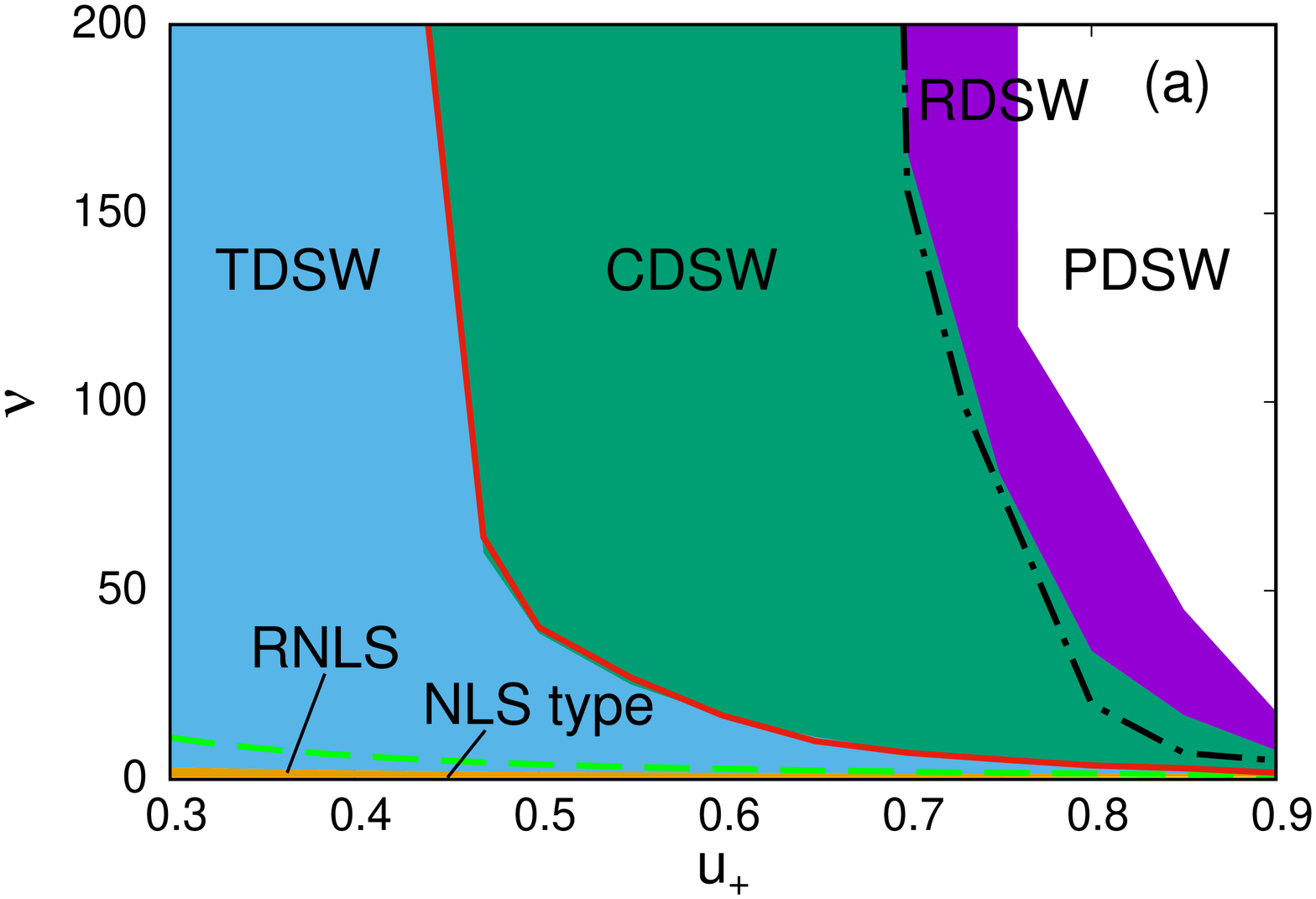}
 \includegraphics[width=0.5\textwidth,angle=270]{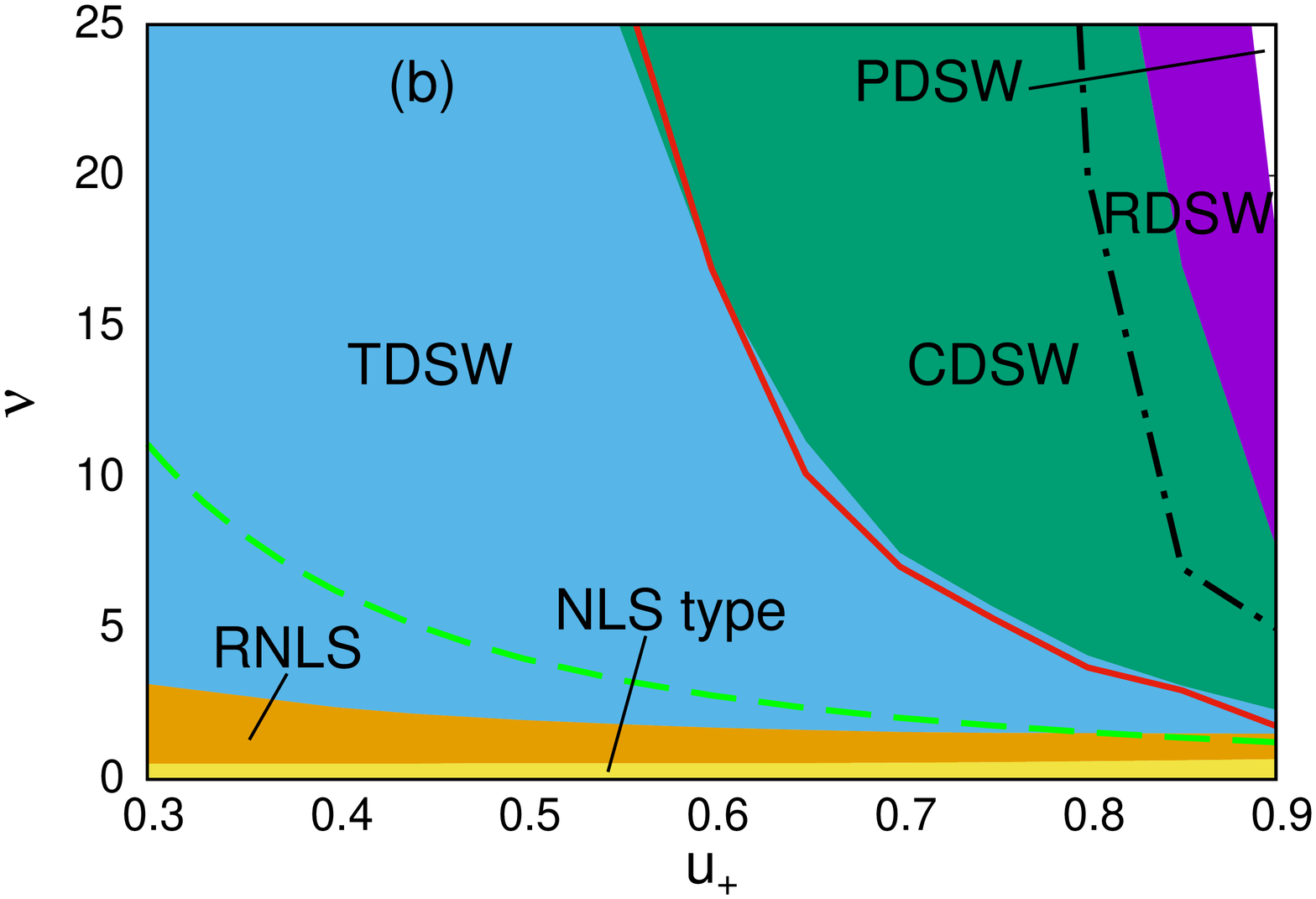}
\caption{Numerical existence regions for nematic DSW types in terms of the nonlocality parameter $\nu$ as the initial level $u_{+}$ varies.  (a) Full existence regions from highly nonlocal $\nu=200$ to local $\nu=0$, (b) Detail of (a) for the transition to an NLS-type DSW. Theoretical boundary (\ref{e:higherphasebore}) between RDSW and CDSW regimes:  black (dot-dash) line; theoretical boundary
$U_{s} = c_{g}$, (\ref{e:lingroup}), between CDSW and TDSW regimes:  red (solid) line;
theoretical boundary (\ref{e:nlsbound}) between KdV-type and NLS-type DSW:  green (dashed) line.  Here $u_{-}=1.0$ and $q=2$.}
\label{f:regions}
\end{figure}

The form and details of the DSW solution of the defocusing nematic equations (\ref{e:eeqnd}) and (\ref{e:direqnd}) will be found as the nonlocality
$\nu$ changes from $O(100)$ to $0$, that is from the nematic having a highly nonlocal response to a local response as the optical power increases.  For high nonlocality, the DSW is of KdV type, with the leading edge of the DSW
consisting of solitary waves of elevation \cite{nembore,nemgennady,saleh}.  The reason for this can be seen from the
eKdV equation (\ref{Pofq}) as for $\nu > q^{2}/(4u_{+}^{2})$ the coefficient of the third derivative is positive,
so that the DSW is of KdV type and has positive polarity.  For $0 \le \nu < q^{2}/(4u_{+}^{2})$ the sign of the third
derivative is negative and the DSW has negative polarity, with solitary waves at the trailing edge and linear dispersive
waves at the leading edge, as for the NLS DSW \cite{gennady}.
A nematic DSW of NLS type is illustrated in Figure
\ref{f:types}(e) for $\nu = 0.5$.  Indeed, for $\nu = 0$, the nematic equations (\ref{e:eeqnd}) and (\ref{e:direqnd}) reduce
to the standard NLS equation on substituting for $\phi$ from (\ref{e:direqnd}) into (\ref{e:eeqnd}).  The DSW solution of
the NLS equation is well known \cite{gennady} as the NLS equation is integrable and so the solution is completely determined.  Hence, the solution for this regime will not be considered here.

The existence regions for the various nematic DSW types as the nonlocality parameter $\nu$ varies (optical power varies),
found from full numerical solutions of the nematic equations (\ref{e:eeqnd}) and (\ref{e:direqnd}) are shown in Figure \ref{f:regions} as the initial level ahead $u_{+}$ varies.  The range $0.3 \le u_{+} \le 0.9$ was chosen
as this encompassed all six of the DSW types studied here.  In addition,
most of the theoretical expressions for the boundaries between these regions and the solutions with each
region were based on $u_{-} - u_{+}$ small, for example the boundary (\ref{e:nlsbound}), which is based on the eKdV equation (\ref{Pofq}).  It can be seen that over most of the $(u_{+},\nu)$ domain the nematic DSW is of CDSW or TDSW type, so that
it is typically unstable.  In addition, it is deduced that
the nematic DSW is of NLS-type only for small values of the nonlocality parameter $\nu$ of $2$ and below.  The nematic DSW is then nonlocal, except for high enough optical powers for which $\nu$ is small.  The nonlocality parameter $\nu$ is given by (\ref{e:nu}).  For the nematic liquid crystal  4-(trans-4-n-hexylcyclohexyl)-isothiocyanato-benzene (6CHBT) the parameter values are $K \sim 10^{-11}N$, $n_{\parallel} = 1.6335$ and $n_{\perp}=1.4967$ \cite{curvenature}.  Let us take the pre-tilt angle $\theta_{0}$ to be $\pi/4$ so that the nematic response is maximised \cite{PR}.  A typical beam wavelength is $1064nm$ and a typical half width $W_{b}$ is $1.5\mu m$ \cite{curvenature}.  With these parameter values, it is found that $\nu=2$ when the beam power $P_{b}$ is
$288 mW$, far in excess of typical beam powers of a few milliWatts to a few tens of milliWatts \cite{PR,curvenature}.  Such a large optical power can result in the nematic medium being heated enough so that its temperature goes above the critical temperature, $43^{o}C$ for 6CHBT \cite{curvenature}, so that it undergoes a phase change out of the nematic state.  In this regard,
it should be noted that experimental nematic cells are small, of the order of $1mm$ in the down cell propagation direction of the beam and $100\mu m\times 10mm$ in cross-section.  The thin cross-section is the direction in which the pre-tilting electric field is applied, which results in a stable and uniform molecular pre-tilt.  We then deduce that for experimental beam powers, the nematic bore will be in the nonlocal response regime with $\nu$ large, that is low optical power.

\section{PDSW and RDSW regimes}
\label{s:pdswrdsw}

Typical PDSW and RDSW solutions are illustrated in Figure \ref{f:types}(a) and (b).  In the PDSW regime the DSW is not in resonance with linear diffractive radiation, so that the DSW is of KdV type, see the boundary (\ref{e:nlsbound}), and is a perturbed KdV-type DSW.  In the PDSW regime the DSW is in resonance with diffractive radiation, so that it consists of a KdV-type DSW with resonant radiation propagating ahead of it.  This resonant radiation is not large enough, however, as to destroy the KdV-type DSW structure, as in the CDSW regime, see Figure \ref{f:types}(c).  As the DSW in the PDSW and RDSW regimes are perturbed KdV DSWs \cite{saleh}, the solutions in these two regimes can be found using the perturbed KdV DSW solution of \cite{perturbkdv}.  In this previous work, the general eKdV equation, a particular case of which is (\ref{Pofq}), was asymptotically transformed to
the KdV equation, whose known DSW solution \cite{elreview,gur} was then used to find the asymptotic DSW solution of the original eKdV equation.  This asymptotic DSW solution can be used here based on the
eKdV equation reduction (\ref{Pofq}) of the full nematic equations in the limit $u_{-} - u_{+}$ small.  The work \cite{perturbkdv} then gives the following PDSW and RDSW nematic DSW solutions. The amplitude of the DSW is
\begin{multline}
    a=2m(u_{i}-u_{+}) + \frac{1}{3}(u_{i}-u_{+})^2\left\{ m\left(1 -m\right)C_1 + m\left(m-2\right)C_2 \right.\\ \left. \mbox{} + mC_3 + 2m\left(8-3m\right)C_4 \right\},
    \label{e:dswamp}
\end{multline}
its wavenumber is
\begin{multline}
    k=\frac{\pi \sqrt{2\left\{u_{i} - u_{+}\right\}}}{K(m)\sqrt{u_{+}}\sqrt{\frac{\nu}{q} - \frac{q}{4u_{+}^{2}}}} \left\{1+\frac{(u_i-u_+)}{12}C_{1} + \frac{(u_i-u_+) (4m^2-8m+3)}{12}C_3 \right.\\ \left.
    \mbox{} - \frac{(u_i-u_+) (8m^2-14m+11)}{3}C_4 \right\},
\end{multline}
and its mean level is
\begin{multline}
    \bar{|u|}= 2u_{+} - u_{i} + (u_{i}-u_{+})\left\{m + \frac{2E(m)}{K(m)} \right\} \\ \left.
    \mbox{} - (u_i-u_+)^2C_{1}\left\{ \frac{3m^2K(m)+4mE(m)-5mK(m)-2E(m)+2K(m)}{18K(m)} \right\} \right. \\
    \mbox{} -(u_i-u_+)^2C_3\left\{\frac{(2-m-m^2)K^2(m)+(4m-10)E(m)K(m) + 8E^2(m))}{6K^2(m)} \right\} \\
    \mbox{} + 2(u_i-u_+)^2C_4\left\{
    \frac{(m^2-7m+6)K^2(m)+2(6m-11)E(m)K(m)+16E^2(m)}{3K^2(m)}\right\}.
    \label{e:meanbore}
\end{multline}
Here, $K(m)$ and $E(m)$ are complete elliptic integrals of the first and second kinds of modulus squared $m$, respectively.  The modulus squared $m$
is a parameter in these amplitude, wavelength and mean height expressions.
It is determined in terms of the simple wave (similarity) variable $x/z$ by
\begin{multline}
    \frac{x}{z} = \frac{\sqrt{2}}{\sqrt{q}}u_{+} + \frac{\sqrt{2}(u_i-u_+)}{\sqrt{q}}\left\{1+m - \frac{2m(1-m)K(m)}{E(m) + (m-1)K(m)}\right\} \\
    \mbox{} + \frac{(u_i-u_+)^2 }{3\sqrt{2q}}C_{1} \left\{1+m - \frac{2m(1-m)K(m)}{E(m) + (m-1)K(m)}\right\} \\
    \mbox{} + \frac{(u_i-u_+)^2}{3\sqrt{2q}} C_{3} \left\{ 2m-1-m^2 - \left(1+m - \frac{2m(1-m)K(m)}{E(m) + (m-1)K(m)}\right) \right.  \\ \left.
    \mbox{} -\frac{4m(1-m)(2E(m)+(m-1)K(m))}{E(m) + (m-1)K(m)} \right\} + \frac{4(u_i-u_+)^2}{3\sqrt{2q}} C_{4} \\
    \times \left\{ -1-m + \frac{2m(1-m)K(m)}{E(m) + (m-1)K(m)} + \left(1+m - \frac{2m(1-m)K(m)}{E(m) + (m-1)K(m)}\right)^2 \right.\\ \left.
    -2(2m-m^2-1) + 4 \frac{m(1-m)[(m-1)K(m)+2E(m)]}{E(m) + (m-1)K(m)}
    \right\} .
    \label{charac}
\end{multline}
This expression for $x/z$ derives from the characteristic of the KdV modulation equations on which the simple wave DSW solution occurs \cite{gur,perturbkdv,bengt}.  The coefficients $C_{j} (j=1,2,3,4)$ in the above solutions are connected to $c_{j} (j=1,2,3,4)$ through the relations
\begin{equation}
  C_{1}=\sqrt{2q}c_{1},\quad C_{2}=\frac{8\sqrt{2}q^{3/2}u_{+}}{(4u^2_+\nu-q^2)}c_{2},\quad C_{3}=\frac{8\sqrt{2}q^{3/2}u_{+}}{(4u^2_+\nu-q^2)}c_{3},\quad C_{4}=\frac{64\sqrt{2}q^{5/2}u^2_+}{(4u^2_+\nu-q^2)^2}c_{4}.
  \label{e:Cs}
\end{equation}
At the leading, solitary wave edge of the DSW $m\to{1}$ and at the trailing, harmonic wave edge of the DSW $m\to{0}$.  It can then be found from the
characteristics (\ref{charac}) that the DSW lies in the range
\begin{multline}
    s_{i}=\sqrt{\frac{2}{q}}\left\{4u_{+}-3u_{i} - (u_i-u_+)^2\left(\frac{1}{2}C_1+C_3-\frac{64}{3}C_4\right)\right\} \\ \le
 \frac{x}{z} \le \sqrt{\frac{2}{q}}\left\{2u_i - u_+ + \frac{1}{3}(u_i-u_+)^2\left(C_1-C_3+4C_4\right)\right\}=s_{+},
\end{multline}
where $s_{i}$ and $s_{+}$ are the harmonic and solitary wave edge velocities of the DSW, respectively.

\begin{figure}
\centering
 \includegraphics[width=0.5\textwidth,angle=270]{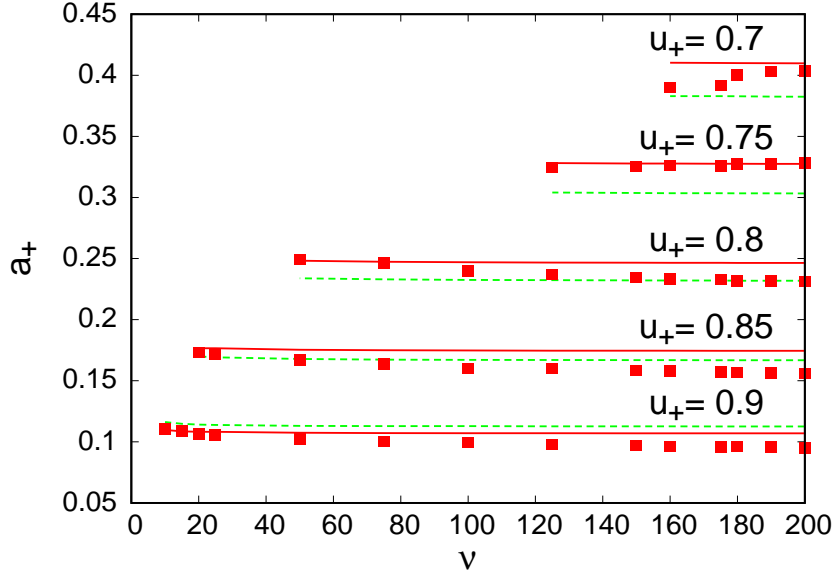}
\caption{Comparison between lead wave amplitude $a_{+}$ of PDSW and RSDW with numerical solutions of the nematic equations (\ref{e:eeqnd}) and (\ref{e:direqnd}) with the initial condition (\ref{e:ic}).  Numerical amplitude:  red squares; amplitude (\ref{e:dswamp}) with $m=1$ given by undular bore solution
of eKdV equation (\ref{Pofq}):  red (solid) line; amplitude given by undular bore solution of Kawahara equation (equation (\ref{Pofq}) with $c_{1}=c_{2}=c_{3}=0$):  green (dashed) line. Here $u_{-}=1.0$ and $q=2$.}
\label{f:pdswrdswamp}
\end{figure}

Comparisons between the lead solitary wave amplitude $a_{+}$ as given by the asymptotic DSW solution (\ref{e:dswamp}), with $m=1$, and numerical solutions
are given in Figure \ref{f:pdswrdswamp} as the nonlocality parameter $\nu$ varies.  The existence regions for the PDSW and RDSW types depend on both $u_{+}$ and $\nu$, see Figure \ref{f:regions}, so that the comparison curves for each $u_{+}$ were stopped at the boundary between the RDSW and
CDSW regimes.  Figure \ref{f:pdswrdswamp} shows the lead
solitary wave amplitude for the full eKdV equation (\ref{Pofq}) and the Kawahara equation, which is (\ref{Pofq}) with $c_{1}=c_{2}=c_{3}=0$.  Previous work on the nematic bore \cite{nemgennady,saleh} was based on the
Kawahara equation, that is only the higher order fifth derivative was included in the asymptotic eKdV equation, so the lead wave amplitude based on this equation is given in the Figure to determine the effect of the extra higher order terms in the full eKdV equation (\ref{Pofq}).
A key observation is that the height of the lead wave of the DSW depends very weakly on the strength of the nonlocality, with little variation even down to $\nu = O(10)$ from the high nonlocality amplitudes with $\nu = O(100)$.  In some sense, the nematic DSW is then nonlocal down to small values of $\nu$ in the PDSW and RDSW regimes, which was also deduced above from Figure \ref{f:regions}.  It can also be seen that the inclusion of the extra higher order terms in the eKdV equation (\ref{Pofq}) over the Kawahara equation improves the agreement with numerical solutions on the whole, especially as the nonlocality parameter $\nu$ decreases, but the effect of these extra terms is small, with the Kawahara equation
giving good agreement over the whole range of $\nu$ and for all values of $u_{+}$, except near the RDSW/CDSW borderline at $u_{+}=0.7$.  This is expected as the weak dependence on the nonlocality parameter $\nu$ means that the DSW is nonlocal, so that $\nu$ can be taken as large.  The dominant higher order term in the eKdV equation is $\varepsilon^{2} c_{4} Q_{\xi\xi\xi\xi\xi}$ in this limit, as noted in (\ref{e:kdv5nem}).  As $u_{+}=0.7$ is approached the resonant radiation shed by the DSW is of relatively large amplitude as in
this limit the RDSW/CDSW boundary is approached.  This results in oscillations in the lead wave
amplitude as the resonant radiation moves through the DSW and is shed.  In these cases, the numerical amplitude shown
in Figure \ref{f:pdswrdswamp} was calculated as an average in $z$ over the last few amplitude oscillations in the numerical solution.  It is noted that except for $u_{+}=0.75$ and $0.7$ the amplitude grows as $\nu$ decreases.
This change in behaviour is due to the DSW changing form as it transitions from the RDSW to the CDSW regime, for which the resonant radiation has a major effect on the DSW with its amplitude decreasing markedly due to the large amount of mass being shed as resonant radiation \cite{saleh}.

The full linear dispersion relation (\ref{e:disp}) for the nematic equations (\ref{e:eeqnd}) and (\ref{e:direqnd}) is
\begin{equation}
 \omega = k\bar{v} + \frac{\sqrt{\bar{\rho}} k}{\sqrt{\nu k^{2} + 2q}}
 \left[ \frac{\nu k^{2} + 2q}{4\bar{\rho}} k^{2} + 4 \right]^{1/2} + \frac{2\bar{\rho}}{q}
 \label{e:dispfull}
\end{equation}
when the mean $2\bar{\rho}/q$ is added.  This mean term arises on
integrating $v = \psi_{x}$ to obtain the dispersion relation for $\psi$
from that for $v$ \cite{nemgennady}.
In the limit $\nu k^{2}$ large, this dispersion relation becomes (\ref{e:displargenu})
\begin{equation}
 \omega = k\bar{v} + \frac{1}{2} k^{2} + \frac{2\bar{\rho}}{q} + \ldots ,
 \label{e:disnularge}
\end{equation}
again with the carrier wave phase shift $2\bar{\rho}/q$ added to obtain
the dispersion relation for $\psi$.
In the RDSW regime the resonant wavetrain has low amplitude, see Figure \ref{f:types}(b).  In addition, this wavetrain has high frequency relative
to the DSW, again see Figure \ref{f:types}(b).  So the appropriate dispersion
relation for the resonant wavetrain is (\ref{e:disnularge}).
In the RDSW regime, the resonance condition for the resonant wavetrain ahead of the DSW has been set by matching the phase velocity of the resonant wavetrain to the velocity of the lead wave of the DSW \cite{nembore,nemgennady,saleh}, giving
\begin{equation}
 s_{+} = c_{r}, \quad c = \bar{v} + \frac{1}{2} k + \frac{2\bar{\rho}}{qk},
 \label{e:rescondition}
\end{equation}
so that
\begin{equation}
 k_{r} = s_{+} + \left[ s_{+}^{2} - \frac{4}{q} u_{+}^{2} \right]^{1/2}
 \label{e:krres}
\end{equation}
on setting $\bar{\rho} = \sqrt{u_{+}}$ and $\bar{v}_{+}=0$ as the resonant
wavetrain propagates on the level ahead.  This gives the wavenumber of
the resonant wavetrain based on this criterion.  The resonant wavetrain
then exists if $s_{+} \ge 2u_{+}/\sqrt{q}$, so that the borderline between
the PDSW and RDSW regimes is $s_{+} = 2u_{+}/\sqrt{q}$.  Previous work
\cite{nemgennady,saleh} has shown that this theoretical borderline is
in excellent agreement with numerical solutions in the high nonlocality
limit $\nu$ large.  For fixed $u_{+}$, as $\nu$ decreases a PDSW changes
to an RDSW, then to a CDSW, see Figure \ref{f:regions}.  For instance, for
$u_{+} = 0.8$, the DSW changes from PDSW to RDSW at $\nu=88$, then to CDSW
at $\nu=34$.  However, the resonance condition (\ref{e:rescondition}),
or (\ref{e:krres}), gives that the DSW changes from PDSW to RDSW at
$\nu = 3$, which is the TDSW regime according to Table \ref{t:typebounds}.  This resonance condition is based on the limit $\nu k^{2} \gg 1$, but even if the full dispersion relation (\ref{e:dispfull}) is used for the
resonance condition (\ref{e:rescondition}) the predicted PDSW/RDSW borderline is $\nu = 2.38$, which is still far from the numerical value
and close to that for $\nu k^{2} \gg 1$.  The resonance condition (\ref{e:rescondition}) is based on resonance between the lead wave of the DSW and diffractive radiation.  However, as pointed out previously \cite{saleh}, a DSW is modulated periodic wave so that all waves of the DSW can resonant with diffractive radiation, not just the lead wave, as seen in Figure \ref{f:type1detail} for a PDSW.  Internal resonance will be discussed in detail in Section \ref{s:cdsw}.  The phase velocity of a component wave of the DSW is (\ref{e:higherphasebore}).  Equating
this bore component phase velocity with the nematic diffractive radiation phase velocity determined from the dispersion relation (\ref{e:disp}) determines the internal resonance.  However, even
using this internal resonance does not result in a borderline between the PDSW and RDSW regimes in any reasonable accord with numerical solutions, see Figure \ref{f:regions}.  A resonant wavetrain will
then exist if the internally resonant waves can propagate out of the DSW, that is their group velocity is greater than the velocity of the lead solitary wave of the DSW.  However, even this condition does not
give the correct boundary between the PDSW and RDSW regimes as the nonlocality parameter $\nu$ decreases.  The issue of internal resonance and its relation to the existence of the PDSW and RDSW regimes merits further study.  In this regard, the recent work \cite{congy} on the interaction of linear wavepackets and DSWs could be relevant.


\section{CDSW regime}
\label{s:cdsw}

For fixed a nonlocality parameter $\nu$, as $u_{+}$ decreases, the jump
height $u_{-} - u_{+}$ increases, the nematic DSW changes from RDSW to CDSW form, see Figures \ref{f:regions}(b) and (c).  The reason for this is that
as the amplitude of the resonant wavetrain grows, it takes more conserved
quantities from the DSW, so that its amplitude is reduced.  In addition, the
DSW becomes unstable in the CDSW regime, as for the Kahawara equation DSW \cite{patkdv}, noting that in the small jump height limit the nematic equations reduce to the eKdV equation (\ref{Pofq}), which becomes the Kahawara equation (\ref{e:kdv5nem}) in the limit of large nonlocality
$\nu$.  Figure \ref{f:types}(c) shows a typical nematic DSW in the CDSW regime.  It can be seen that the DSW has changed from a modulated wavetrain with a monotonically decreasing amplitude from front to rear to a disordered wavetrain with an essentially uniform amplitude on average,
except at its rear.  This structure is in agreement with unstable DSW structure for the focussing NLS equation \cite{tovbis}.  The
unstable DSW can then be approximated by a train of equal amplitude solitary waves, which has been found to give good results for DSW solutions \cite{boreapprox}, particularly unstable DSWs, and for the particular case of the nematic CDSW in the high nonlocality, low optical power, regime \cite{saleh}.  The amplitude of the solitary waves of the CDSW is determined from mass and energy conservation equations for the underlying nonlinear dispersive wave equation \cite{boreapprox}.

If we set
\begin{equation}
 B_{2} = 3 \sqrt{\frac{2}{q}}, \quad B_{3} = \frac{4\nu u_{+}^{2}
 - q^{2}}{8\sqrt{2}q^{3/2}u_{+}}
 \label{e:b1b2}
\end{equation}
for simplicity, the eKdV equation (\ref{Pofq}) has the mass conservation equation
\begin{eqnarray}
& & \frac{\partial}{\partial \eta} Q + \frac{\partial}{\partial \xi}
 \left[ \frac{1}{2}B_{2} Q^{2} + B_{3}Q_{\xi\xi} + \frac{1}{3}\varepsilon^{2} c_{1}Q^{3} + \frac{1}{2}\varepsilon^{2} \left( c_{2} - c_{3} \right) Q_{\xi}^{2} + \varepsilon^{2} c_{3} QQ_{\xi\xi} \right. \nonumber \\
 & & \left. \mbox{} + \varepsilon^{2} c_{4} Q_{\xi\xi\xi\xi} \right] = 0.
 \label{e:masscons}
\end{eqnarray}
The derivation of the energy conservation for the eKdV equation (\ref{Pofq}) is not as straightforward.  Multiplying the eKdV equation by $Q$ and integrating by parts gives
\begin{eqnarray}
 & & \frac{\partial}{\partial \eta} \frac{1}{2}Q^{2} + \frac{\partial}{\partial \xi} \left[ \frac{1}{3}B_{2}Q^{3} + B_{3}QQ_{\xi\xi}
 - \frac{1}{2}B_{3} Q_{\xi}^{2} + \frac{1}{4} \varepsilon^{2} c_{1} Q^{4}
 + \frac{1}{2} \varepsilon^{2} c_{2}Q^{2}Q_{\xi\xi} \right. \nonumber \\
 & & \left. \mbox{} + \varepsilon^{2} c_{4}QQ_{\xi\xi\xi\xi}
 - \varepsilon^{2} c_{4}Q_{\xi}Q_{\xi\xi\xi} + \frac{1}{2} \varepsilon^{2} c_{4} Q_{\xi\xi}^{2} \right] + \varepsilon^{2} \left( c_{3} - \frac{1}{2}c_{2} \right) Q^{2}Q_{\xi\xi\xi} = 0.
 \label{e:energycons}
\end{eqnarray}
The final term on the right had side of this equation cannot be expressed as a perfect derivative.  However, it can be approximately expressed in this form on noting that $\varepsilon$ is small, so that at first order the eKdV equation (\ref{Pofq}) is the KdV equation
\begin{equation}
 \frac{\partial Q}{\partial \eta} + B_{2}Q\frac{\partial Q}{\partial \xi}
 + B_{3} \frac{\partial^{3}Q}{\partial \xi^{3}} = 0.
 \label{e:kdv}
\end{equation}
We then have at leading order that
\begin{equation}
 \frac{\partial}{\partial \eta} Q^{3} = -3Q^{2} \left( B_{2} QQ_{\xi}
 + B_{3}Q_{\xi\xi\xi} \right) = -\frac{\partial}{\partial \xi} \frac{3}{4}B_{2}Q^{4} - 3B_{3}Q^{2}Q_{\xi\xi\xi}.
 \label{e:q3}
\end{equation}
This expression may now be used to eliminate the term in $QQ_{\xi\xi\xi}$ in equation (\ref{e:energycons}) to give the final energy conservation equation
\begin{eqnarray}
 & & \frac{\partial}{\partial \eta} \left[ \frac{1}{2}Q^{2} - \varepsilon^{2}
  \frac{c_{3} - \frac{1}{2}c_{2}}{3B_{3}} Q^{3} \right] + \frac{\partial}{\partial \xi} \left[ \frac{1}{3}B_{2}Q^{3} + B_{3}QQ_{\xi\xi}
 - \frac{1}{2}B_{3} Q_{\xi}^{2} + \frac{1}{4} \varepsilon^{2} c_{1} Q^{4}
 \right. \nonumber \\
 & & \left. \mbox{} + \frac{1}{2} \varepsilon^{2} c_{2}Q^{2}Q_{\xi\xi}
 + \varepsilon^{2} c_{4}QQ_{\xi\xi\xi\xi}
 - \varepsilon^{2} c_{4}Q_{\xi}Q_{\xi\xi\xi} + \frac{1}{2} \varepsilon^{2} c_{4} Q_{\xi\xi}^{2} \right. \nonumber \\
 & & \left. \mbox{} - \varepsilon^{2} \frac{B_{2}}{4B_{3}} \left( c_{3} - \frac{1}{2} c_{2} \right) Q^{4} \right] = 0,
 \label{e:energyconsf}
\end{eqnarray}
which is accurate to $O(\varepsilon^{2})$.

To obtain an approximation to the nematic CDSW, let us assume that at position $\eta$ the CDSW consists of $N$ equal solitary waves of
amplitude $\tilde{a}_{s}$ and width $\tilde{w}_{s}$ \cite{boreapprox}, where we shall use tildes to denote scaled variables in the moving
and stretched coordinates $(\xi,\eta)$.  It is also assumed that the CDSW sheds a uniform downstream resonant wavetrain of (scaled)
amplitude $\tilde{a}_{r}$.  Then as $\xi \to -\infty$, $Q \to 1$ and as
$\xi \to \infty$, $Q \to \tilde{a}_{r} \cos \left( \tilde{k}_{r} \xi - \tilde{\omega}_{r}\eta \right)$, since
$|u| = u_{+} + \varepsilon^{2} Q$ with $\varepsilon^{2} = u_{i} - u_{+}$.
As the CDSW is approximated by a train of solitary waves, the solitary wave solution of the eKdV equation (\ref{Pofq}) is also needed.  While there is no known exact solitary wave solution of this equation, there is an asymptotic solution for $\varepsilon \ll 1$ \cite{timsol}.  To use this solution, the eKdV equation (\ref{Pofq}) needs to be rescaled to conform with the eKdV scaling of \cite{timsol}.  Performing this, we find that the asymptotic solitary wave solution of the eKdV equation (\ref{Pofq}) is
\begin{equation}
 Q = \gamma_{1} \sech^{2} \frac{\xi - V_{s}\eta}{W} +
 \gamma_{2} \sech^{4} \frac{\xi - V_{s} \eta}{W}, \quad
 V_{s} = \frac{1}{3} B_{2}A\left( 1 + 2 \varepsilon^{2} C_{4}A \right) ,
 \label{e:hsol}
\end{equation}
where
\begin{equation}
 W = \frac{\sqrt{12B_{3}}}{\sqrt{B_{2}A}}, \quad \gamma_{1} = A + \varepsilon^{2}C_{6}A^{2}, \quad  \gamma_{2} = \varepsilon^{2} C_{7} A^{2}.
\label{e:gamma1gamma2}
\end{equation}
The rescaled coefficients $c_{i},$ $i=1,\ldots,4$, of the eKdV equation
(\ref{Pofq}), denoted by $C_{i}$, $i=1,\ldots,4$, are given by
\begin{eqnarray}
C_{1} = \frac{6}{B_{2}} c_{1}, \quad C_{2} = \frac{1}{B_{3}} c_{2}, & &
C_{3} = \frac{1}{B_{3}} c_{3}, \quad C_{4} = \frac{B_{2}}{6B_{3}^{2}} c_{4}, \nonumber \\
 C_{6} = \frac{2}{3}C_{3} - \frac{1}{6}C_{1} + \frac{1}{6} C_{2} - 5C_{4}, & &  C_{7} = \frac{15}{2}C_{4} - \frac{1}{2}C_{3} + \frac{1}{12}C_{1} - \frac{1}{4} C_{2}.
 \label{e:newc}
\end{eqnarray}
It is noted that these scaled $C_{i}$, $i=1,\ldots,4$, are the same as the $C_{i}$ (\ref{e:Cs}) used for the perturbed DSW solution
(\ref{e:dswamp})--(\ref{charac}) due to the same rescaling from the eKdV equation (\ref{Pofq}) being used for this solution.

\begin{figure}
\centering
 \includegraphics[width=0.5\textwidth,angle=270]{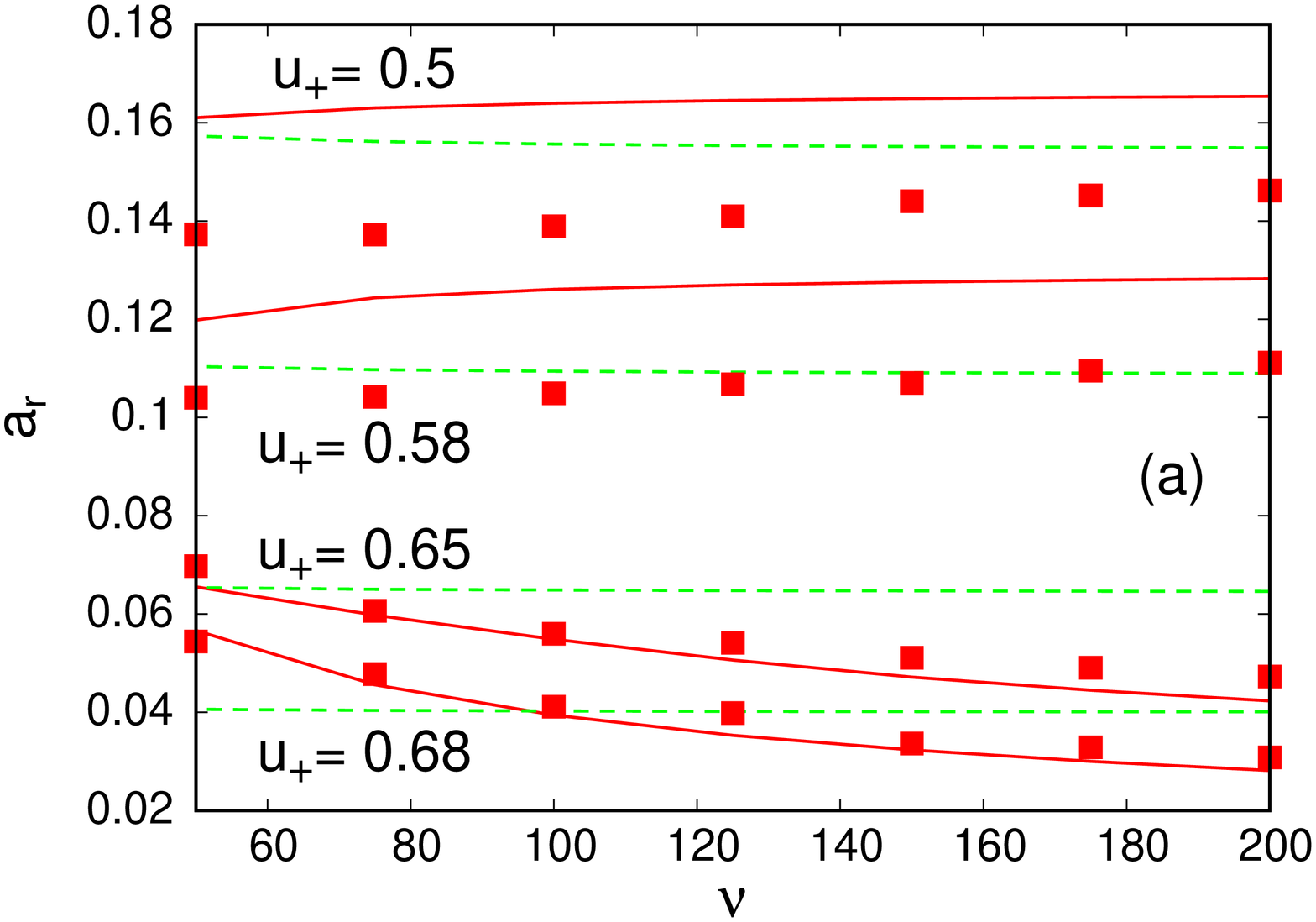}
 \includegraphics[width=0.5\textwidth,angle=270]{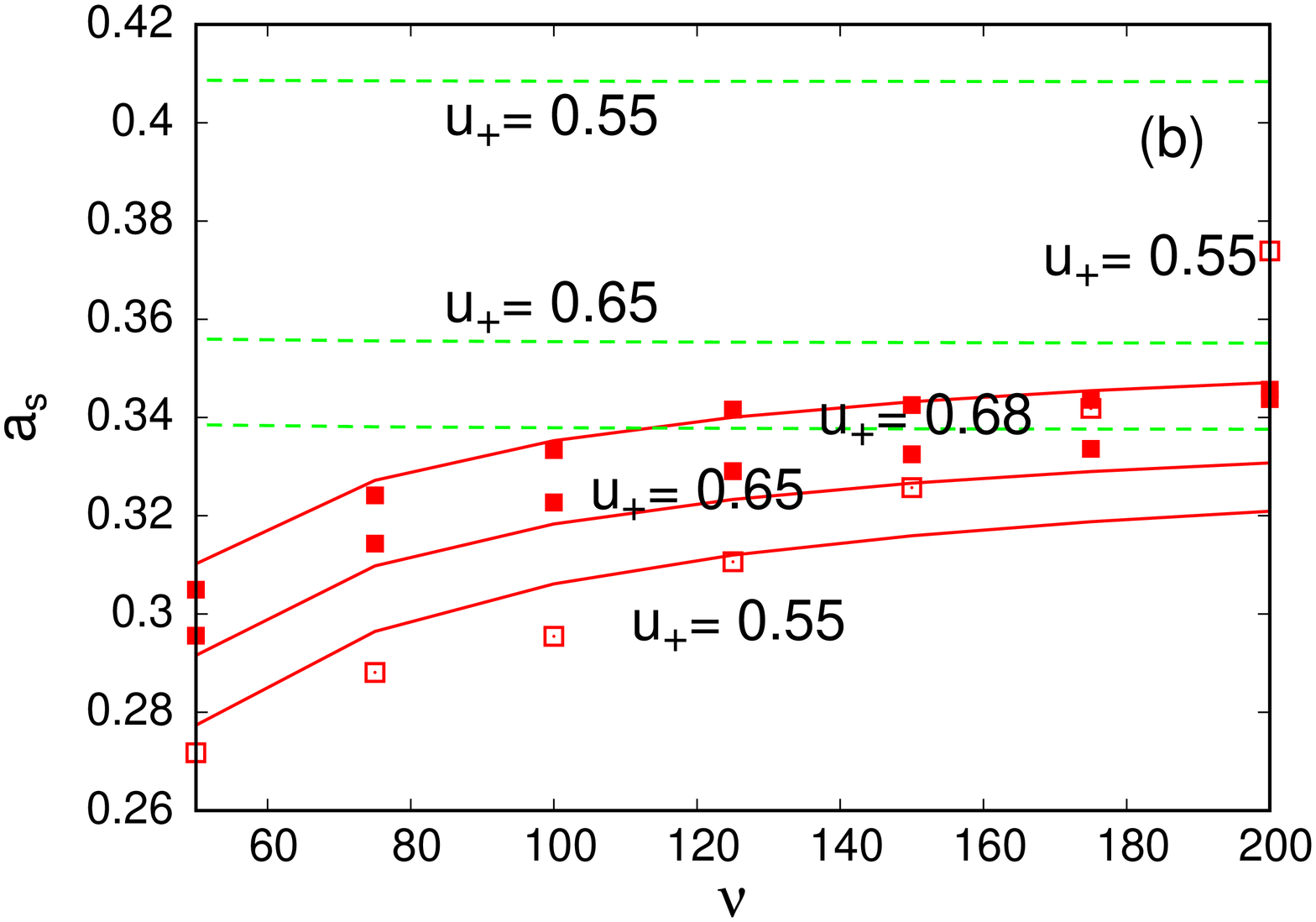}
\caption{Comparison between theory and numerical solutions of the nematic equations (\ref{e:eeqnd}) and (\ref{e:direqnd}) with the initial condition (\ref{e:ic}) for the nematic CDSW.  Numerical amplitude:  red squares; amplitude (\ref{e:ar}):  red (solid) line; amplitude given by the Kawahara equation (equation (\ref{Pofq}) with $c_{1}=c_{2}=c_{3}=0$):  green (dashed) line. (a) Resonant wave amplitude $a_{r}$, (b) CDSW solitary wave amplitude $a_{s}$ (\ref{cdswamp}).  Note that for clarity the numerical DSW amplitude for $u_{+}=0.55$ is denoted by red open squares.  Here $u_{-}=1.0$ and $q=2$.}
\label{f:cdswresamp}
\end{figure}

The mass and energy conservation equations (\ref{e:masscons}) and (\ref{e:energyconsf}) can now be used to determine the amplitude of the solitary waves in the nematic CDSW in the equal amplitude approximation.
The mass and energy conservation equations are integrated in $\xi$ from
$-\infty$ to $\infty$.  The integral of the mass and energy densities are approximated by $N$ times these for a single solitary wave (\ref{e:hsol}).
The flux terms are calculated using the boundary conditions stated above, $Q = 1$ at $\xi = -\infty$ and $Q = \tilde{a}_{r} \cos \left( \tilde{k}_{r} \xi - \tilde{\omega}_{r}\eta \right)$ at $\xi = \infty$.  The
resonant radiation flux at $\xi = \infty$ is calculated by averaging the
periodic radiation over a period \cite{saleh}.  In this manner, integrating the mass conservation equation gives
\begin{equation}
 N \left[ 2\gamma_{1} + \frac{4}{3}\gamma_{2} \right] W
 = \left\{ \frac{1}{2}B_{2} + \frac{1}{3}\varepsilon^{2} c_{1}
 - \left[ \frac{1}{4}B_{2} + \frac{1}{4} \varepsilon^{2}
 \left( c_{2} - 3c_{3} \right) \tilde{k}_{r}^{2}\right] \tilde{a}_{r}^{2}
 \right\} \eta
 \label{e:massloss}
\end{equation}
and integrating the energy conservation equation gives
\begin{eqnarray}
& &  N \left[ \frac{2}{3} \gamma_{1}^{2} + \frac{16}{15}\gamma_{1}\gamma_{2}
  - \varepsilon^{2} \frac{16}{45}\frac{c_{3} - \frac{1}{2}c_{2}}{B_{3}} \gamma_{1}^{3} \right] W
  = \left\{ \frac{1}{3}B_{2} + \frac{1}{4}\varepsilon^{2} c_{1}
  \nonumber \right. \\
  & & \mbox{} - \varepsilon^{2} \frac{B_{2}}{4B_{3}} \left( c_{3} - \frac{1}{2}c_{2} \right) - \frac{1}{4}\left[ -3B_{3}\tilde{k}_{r}^{2}\tilde{a}_{r}^{2}
  + \frac{3}{8} \varepsilon^{2} c_{1} \tilde{a}_{r}^{4} + 5 \varepsilon^{2}
  c_{4} \tilde{k}_{r}^{4}\tilde{a}_{r}^{2} \right.  \nonumber \\
  & & \left. \left. \mbox{} - \varepsilon^{2} \frac{3B_{2}}{8B_{3}}
  \left( c_{3} - \frac{1}{2} c_{2} \right) \tilde{a}_{r}^{4} \right] \right\} \eta \nonumber \\
  & & \sim \left\{ \frac{1}{3}B_{2} + \frac{1}{4}\varepsilon^{2} c_{1}
  - \varepsilon^{2} \frac{B_{2}}{4B_{3}} \left( c_{3} - \frac{1}{2}c_{2} \right) - \frac{1}{4}\tilde{c}_{g} \tilde{a}_{r}^{2} \right\} \eta
  \label{e:energyloss}
\end{eqnarray}
since $\tilde{a}_{r}$ is small.  In addition, this neglect of quartic terms in $\tilde{a}_{r}$ is consistent with the radiation being determined
by a linear WKB analysis \cite{nemgennady}.  Here, $\tilde{c}_{g}$ is the scaled group velocity of the resonant radiation based on (\ref{scales}).  Dividing the mass and energy equations (\ref{e:massloss}) and (\ref{e:energyloss}) gives an
equation for $A$
\begin{equation}
 \frac{\gamma_{1}^{2} + \frac{8}{5}\gamma_{1}\gamma_{2}
  - \varepsilon^{2} \frac{8}{15}\frac{c_{3} - \frac{1}{2}c_{2}}{B_{3}} \gamma_{1}^{3}}{\gamma_{1} + \frac{2}{3}\gamma_{2}} =
  4\frac{B_{2} + \frac{3}{4}\varepsilon^{2} c_{1}
  - \varepsilon^{2} \frac{3B_{2}}{4B_{3}} \left( c_{3} - \frac{1}{2}c_{2} \right) - \frac{3}{4}\tilde{c}_{g} \tilde{a}_{r}^{2}}{2B_{2} + \frac{4}{3}\varepsilon^{2} c_{1}
 - \left[ B_{2} + \varepsilon^{2}
 \left( c_{2} - 3c_{3} \right) \tilde{k}_{r}^{2}\right] \tilde{a}_{r}^{2}}
 \label{e:Aeqn}
\end{equation}
in terms of $\tilde{a}_{r}$.  Once $A$ is determined the unscaled amplitude $a_{s}$ of the solitary waves of the CDSW is given by
\begin{equation}
 a_{s} = \varepsilon^{2} \left[ A + \varepsilon^{2} \left( C_{6} + C_{7}
 \right)A^{2} \right] = (u_{i}-u_{+}) \left[ A + (u_{i}-u_{+}) \left( C_{6} + C_{7} \right)A^{2} \right] ,
 \label{cdswamp}
\end{equation}
on using the solitary wave solution (\ref{e:hsol}).
Transforming back from the scaled eKdV variables to the original variables, the relation (\ref{e:Aeqn}) becomes
\begin{eqnarray}
& &  \frac{\gamma_{1}^{2} + \frac{8}{5}\gamma_{1}\gamma_{2}
  - \frac{8}{15}(u_{i}-u_{+})\frac{c_{3} - \frac{1}{2}c_{2}}{B_{3}} \gamma_{1}^{3}}{\gamma_{1} + \frac{2}{3}\gamma_{2}} \nonumber \\
  & & =
  4\frac{B_{2} + \frac{3}{4}(u_{i}-u_{+}) c_{1}
  - \frac{3B_{2}}{4B_{3}}(u_{i}-u_{+}) \left( c_{3} - \frac{1}{2}c_{2} \right) - \frac{3}{4}\frac{\left(c_{g} - \sqrt{\frac{2}{q}}u_{+} \right) a_{r}^{2}}{(u_{i}-u_{+})^{3}}}{2B_{2} + \frac{4}{3}(u_{i}-u_{+}) c_{1}
 - \left[ B_{2} +
 \left( c_{2} - 3c_{3} \right) k_{r}^{2}\right] \frac{a_{r}^{2}}{(u_{i}-u_{+})^{2}}}
 \label{e:Aeqnorig}
\end{eqnarray}
on using the scalings (\ref{scales}) for the eKdV equation.
Substituting for the higher order solitary wave coefficients $\gamma_{1}$ and $\gamma_{2}$ given by (\ref{e:gamma1gamma2}) gives the final equation determining the amplitude of the CDSW solitary waves as
\begin{eqnarray}
& &  A\frac{1 + 2(u_{i} - u_{+})\left( C_{6} + \frac{4}{5}C_{7} \right) A
- \frac{8}{15}(u_{i}-u_{+})\left( C_{3} - \frac{1}{2}C_{2}\right) A}
{1 + (u_{i} - u_{+})\left( C_{6} + \frac{2}{3}C_{7} \right) A} \nonumber \\
  & & =
  4\frac{B_{2} + \frac{1}{8} \left( u_{i} - u_{+} \right) B_{2} \left( C_{1} + 3C_{2} - 6C_{3} \right) - \frac{3}{4}\frac{\left(c_{g} - \sqrt{\frac{2}{q}}u_{+} \right) a_{r}^{2}}{(u_{i}-u_{+})^{3}}}{2B_{2} + \frac{2}{9}(u_{i}-u_{+}) B_{2}C_{1} - \left[ B_{2} +
 \left( c_{2} - 3c_{3} \right) k_{r}^{2}\right] \frac{a_{r}^{2}}{(u_{i}-u_{+})^{2}}}.
 \label{e:Aeqnorigsimp}
\end{eqnarray}

\begin{figure}
\centering
 \includegraphics[width=0.25\textwidth,angle=270]{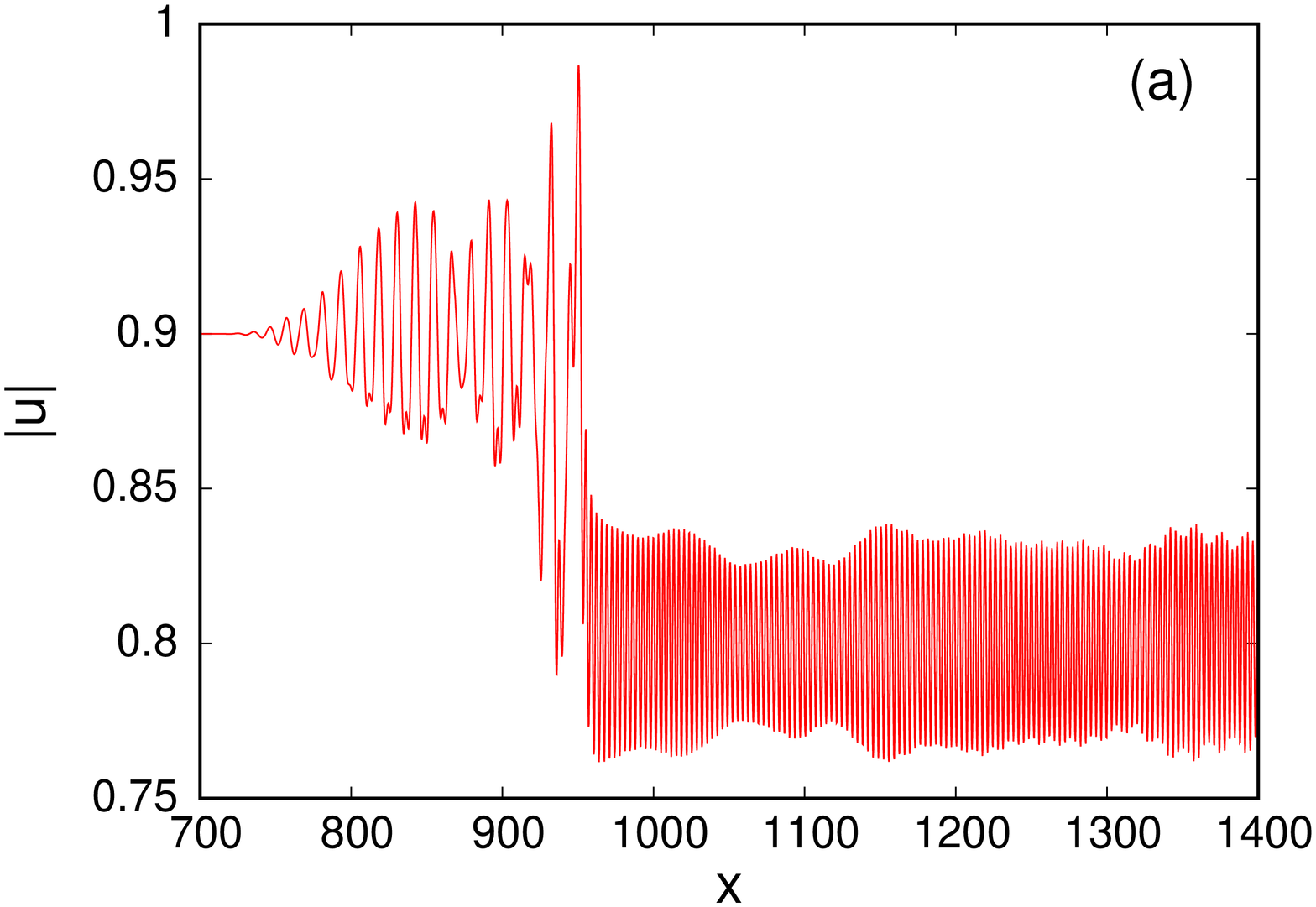}
 \includegraphics[width=0.25\textwidth,angle=270]{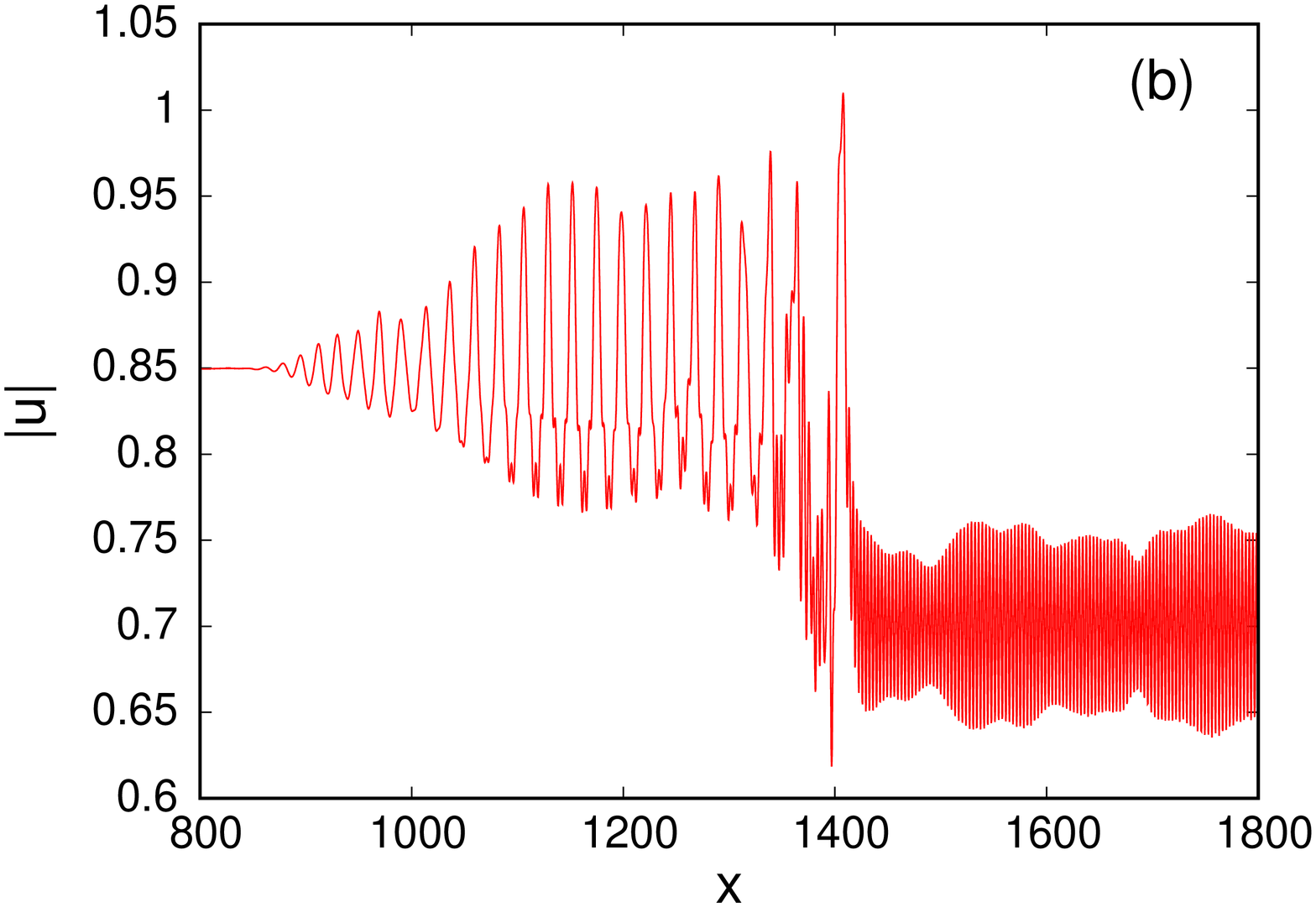}
 \includegraphics[width=0.25\textwidth,angle=270]{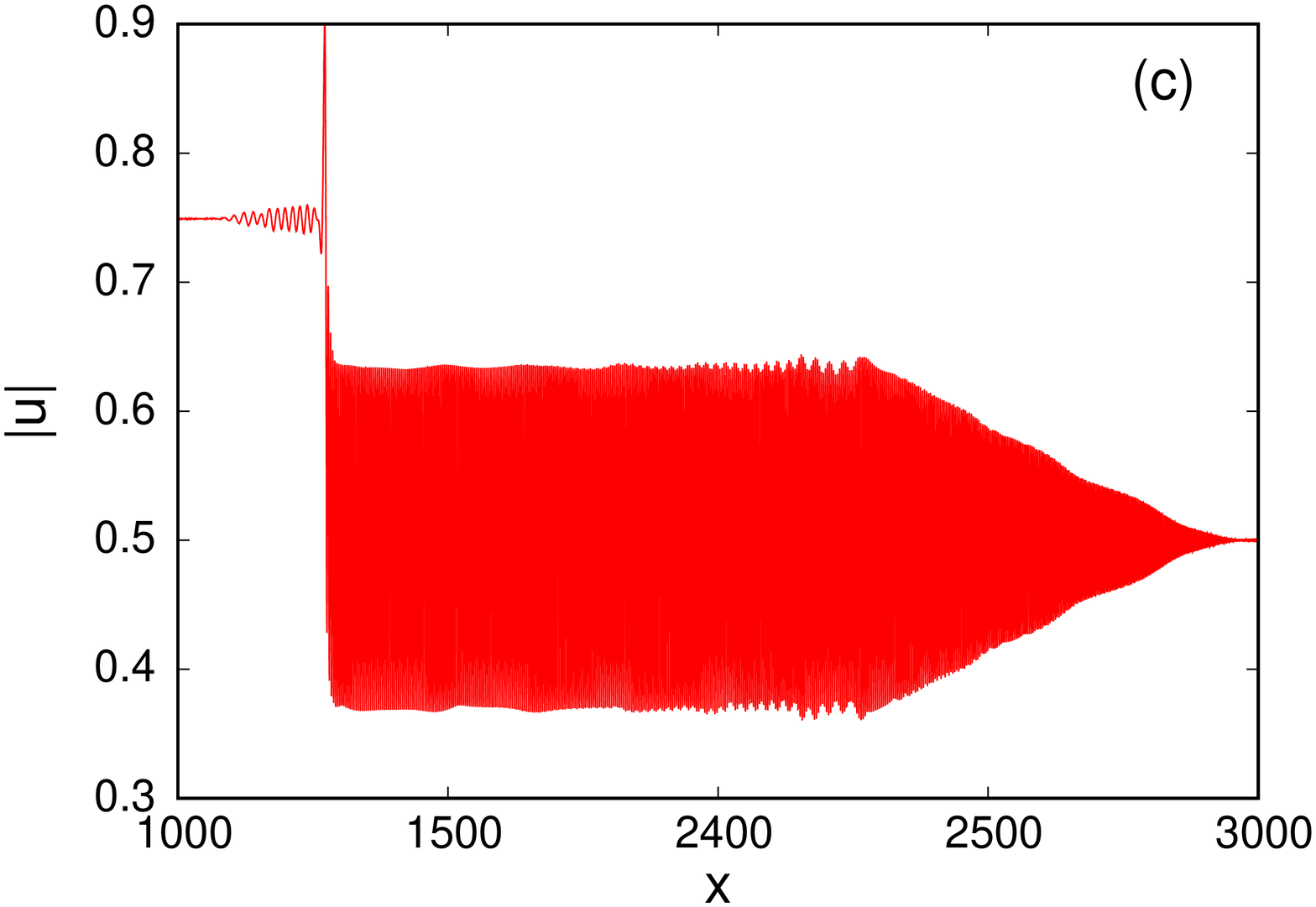}
\caption{Numerical solution of nematic equations (\ref{e:eeqnd}) and (\ref{e:direqnd}) in the CDSW regime for the
initial condition (\ref{e:ic}).  (a) $u_{+}=0.8$ at $z=1000$, (b) $u_{+}=0.7$ at $z=1500$, (c) $u_{+}=0.5$ at $z=1500$.   Here $u_{-}=1.0$ and $q=2$.}
\label{f:type3detail}
\end{figure}

The final quantity to determine is the amplitude of $a_{r}$ of the resonant radiation, which is related to the scaled amplitude $\tilde{a}_{r}$ by $a_{r} = \varepsilon^{2} \tilde{a}_{r} = (u_{i} - u_{+}) \tilde{a}_{r}$.  This resonant radiation was determined as a WKB solution of the nematic equations (\ref{e:eeqnd}) and (\ref{e:direqnd}) by linearising about the mean level $u_{+}$ of the resonant radiation \cite{nemgennady}.
This WKB solution gives the amplitude of the resonant radiation as
\begin{equation}
 a_{r} = \frac{1}{2} \frac{u_{-} - u_{+}}{1 + \frac{2u_{+} k_{r} a_{s}}{qs_{+} (k_{r}-s_{+})^{2}}}.
 \label{e:ar}
\end{equation}
Here, $s_{+}$ is the unscaled velocity of the CDSW, which is \cite{saleh}
\begin{equation}
 s_{+} = \sqrt{\frac{2}{q}} u_{+} + \frac{1}{3}B_{2} a_{s} \left( 1 + 2C_{4}a_{s} \right).
 \label{e:spcdsw}
\end{equation}
The resonant radiation wavenumber $k_{r}$ is determined by the resonance condition (\ref{e:krres}) and the group velocity of the resonant radiation is given by the $k$ derivative of the short wave dispersion relation (\ref{e:disnularge}).  The resonant radiation is a solution of the nematic equations in the limit $\nu k^{2} \gg 1$, so that the appropriate group velocity for it is that from the dispersion relation (\ref{e:disnularge}), not the linearised KdV group velocity of the eKdV equation (\ref{Pofq}) \cite{saleh}.

Figure \ref{f:cdswresamp} displays comparisons of the nematic DSW amplitude $a_{s}$ and resonant wave amplitude $a_{r}$ in the CDSW regime as given by (\ref{e:Aeqnorig}) and (\ref{e:ar}) and numerical solutions.  A typical CDSW is shown in Figure \ref{f:types}(c), with details of the actual CDSW of this figure shown in Figure \ref{f:type3detail}(a).  It can be seen that the lead waves of the DSW have an approximately uniform amplitude, with a rapid decrease of the amplitude towards the trailing edge of the CDSW, as also illustrated in Figure \ref{f:type3detail}(b).  This solution, and that of Figure \ref{f:type3detail}(a), are typical structures for an unstable DSW
\cite{tovbis}.  The numerical DSW amplitude for the comparisons of this figure was calculated as an average over the approximately uniform waves at the leading edge, which is the same assumption on which the equal amplitude approximation used to calculate the solitary wave amplitude $a_{s}$ was
based.  Figure \ref{f:cdswresamp}(a) shows comparisons for the amplitude $a_{r}$ of the resonant wavetrain leading the CDSW, see Figure \ref{f:types}(c).  It can be seen that there is excellent agreement between the theoretical amplitude with the numerical amplitude for the larger values of the level ahead $u_{+}$, with the increase of $a_{r}$ as the nonlocality parameter $\nu$ decreases being correctly given.  This agreement is much improved through the inclusion of the all the higher order terms in the eKdV  equation (\ref{Pofq}) than that of previous work \cite{saleh} based on the Kawahara equation, for which $c_{1}=c_{2}=c_{3}=0$, as given by the green dashed line in the figure.  As $u_{+}$ decreases and the TDSW regime is approached, the agreement between theory and numerical solutions decreases.  This is shown particularly in the final comparison of Figure \ref{f:cdswresamp}(a) for $u_{+} = 0.5$, which is near the TDSW boundary, see Figure \ref{f:regions}.  
The reason for this decreasing agreement is that as the TDSW regime is approached the number of waves in the CDSW decreases so that only one lead wave is left, see Figure \ref{f:type3detail}(c).  The approximation that an average can be taken over an equal amplitude wavetrain then breaks down.

Figure \ref{f:cdswresamp}(b) displays similar comparisons for the amplitude $a_{s}$ of the nematic CDSW with numerical solutions.  It can be seen that the comparison for the DSW amplitude is similar to that for the resonant wave amplitude.  It should be noted that different values of $u_{+}$ have been chosen for the DSW amplitude comparisons for the sake of clarity.
The inclusion of all the higher order terms in the eKdV equation (\ref{Pofq}) results in a significant improvement in the agreement with numerical solutions over that based on the Kawahara equation with $c_{1}=c_{2}=c_{3}=0$ when the level ahead $u_{+}$ is away from the TDSW/CDSW boundary of Figure \ref{f:regions}.  As $u_{+}$ approaches the TDSW/CDSW boundary, 
the DSW amplitude as given by the eKdV equation differs significantly from the numerical amplitude.  The reason for this is that discussed in the previous paragraph for the resonant wave amplitude, the fact that the CDSW ceases to be a train of equal amplitude solitary waves, but reduces to a few waves.

\begin{figure}
\centering
 \includegraphics[width=0.5\textwidth,angle=270]{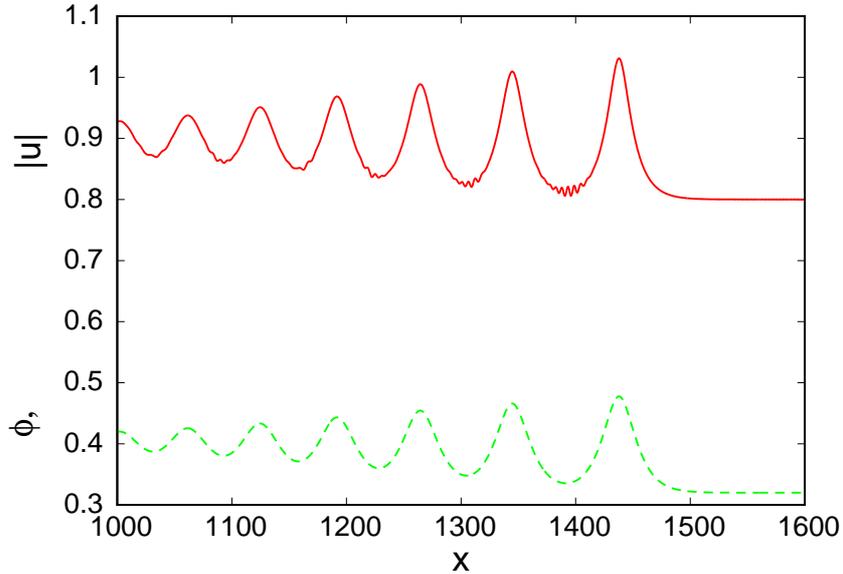}
\caption{Numerical solution of nematic equations (\ref{e:eeqnd}) and (\ref{e:direqnd}) in PDSW regime at $z=1500$ for
initial condition (\ref{e:ic}) with $u_{+} = 0.8$ and $u_{-} = 1.0$.  Red (solid) line:  $|u|$; green (dashed) line: $\phi$.   Here $\nu = 200$ and $q=2$.}
\label{f:type1detail}
\end{figure}

The next thing that we discuss is the analytical borderline between the nematic CDSW and RDSW regimes. This borderline can be found by determining when the resonant amplitude (\ref{e:ar}) approaches zero or approaches a minimum as a function of the nonlocality parameter $\nu$. The determination of this borderline is similar to that of previous work \cite{saleh} in which the resonant amplitude was found as a function of the initial state $u_{+}$, rather than $\nu$. This approach gave a borderline for $u_{+}\leq{0.73}$ for large $\nu$, with the CDSW regime not existing above this value of $u_{+}$ for large $\nu$.  Above this limit, the resonance condition (\ref{e:rescondition}) ceases to work as a function of $\nu$ and an alternative method needs to be found to determine the borderline. To determine the borderline in this case we exploit the fact that the (interior) structure of the nematic DSW is resonant, as evidenced in Figure \ref{f:type1detail}, not only the leading, solitary wave edge.
A DSW is an extended modulated periodic wavetrain, so that all its component waves can resonant, not just the leading edge.  To verify this internal resonance, the phase velocity of the modulated cnoidal waves forming the DSW needs to be matched with the nematic linear phase velocity on the local mean level of the DSW. By way of illustration, we equate the DSW phase velocity \cite{perturbkdv}
\begin{multline}
    c_{p} = \sqrt{\frac{2}{q}}u_{+} + \frac{(u_{i}-u_{+})}{\sqrt{2q}}\left\{ 2+ 2m - \frac{1}{7}(u_{i}-u_{+})(8C_4-C_3)(2m-m^2-1) \right. \\ \left. +  \frac{1}{3}(u_{i}-u_{+})C_{4}(2+2m)^2 - \frac{1}{6}(u_{i}-u_{+})(C_3-C_1+4C_4)(2+2m) \right\},
    \label{e:higherphasebore}
\end{multline}
where $C_1$, $C_3$ and $C_4$ are given in (\ref{e:Cs}), with the nematic phase velocity from the full linear dispersion relation (\ref{e:dispfull}) on the DSW background (\ref{e:meanbore}) and solved for internal resonant wavenumbers, which are always positive and real, as the modulus $m$ varies from near zero to near one.  As the DSW parameter expressions (\ref{e:meanbore}) and (\ref{e:higherphasebore}) are only valid for a well-ranked DSW (stable DSW), such as the PDSW and the RDSW, and the CDSW is an ill-ranked DSW (unstable DSW), then a borderline exists
when these wave parameter expressions result in imaginary internal resonant wavenumbers, which are unphysical, at a borderline value of $\nu$.  The mean flow $\bar{v}$ in the dispersion relation (\ref{e:dispfull}) was determined from the extended KdV reduction of the nematic equations of Section \ref{s:pdswrdsw}.  Substituting the $O(\varepsilon^{2})$ mean flow $v_{1}$ given by (\ref{e:v1}) into the mean flow perturbation
expansion (\ref{e:phikdv}) for $v$ and then averaging gives
\begin{equation}
 \bar{v} = \frac{2\sqrt{2}}{\sqrt{q}} \left( \bar{|u|} - u_{+} \right).
 \label{e:barv}
\end{equation}
This then completes the determination of the CDSW/RDSW borderline.
A comparison between the theoretical and numerical borderlines between the RDSW and CDSW regimes is shown in Figure \ref{f:regions}.
It can be seen that the theoretical borderline is in excellent agreement with the numerical borderline for large values of the nonlocality parameter $\nu$ down to around $\nu=50$, with poorer agreement towards
the local limit with $\nu$ small for which the DSW is changing form from
KdV type to NLS type.

\section{TDSW regime}

Figure \ref{f:types}(d) displays a typical DSW in the TDSW regime.  There is (almost) a constant amplitude resonant wavetrain which at its trailing edge is connected to the intermediate level $u_{i}$.  At its leading edge there is a modulated wavetrain which takes $u$ down to the level $u_{+}$ ahead.
This wavetrain leading the resonant wave is a partial DSW \cite{pat,saleh,resflowmod}.  A partial DSW differs from a standard DSW in that it connects a uniform state to a periodic wavetrain, unlike a standard DSW which links two different levels.  While there is a negative polarity solitary wave connecting the resonant wavetrain to the intermediate level \cite{patkdv}, this connection can be approximated by a Whitham shock \cite{saleh,patjump}, a shock, a jump, in the modulation parameters of the wavetrain, wavelength, frequency, amplitude and mean level, of the Whitham modulation equations for the modulated periodic wavetrain
\cite{whitham}.  A Whitham shock is determined from the Whitham modulation equations.  As noted above, the nematic DSW is in the nonlocal regime, so the appropriate Whitham modulation equations are those for $\nu$ large.

\begin{figure}
\centering
 \includegraphics[width=0.5\textwidth,angle=270]{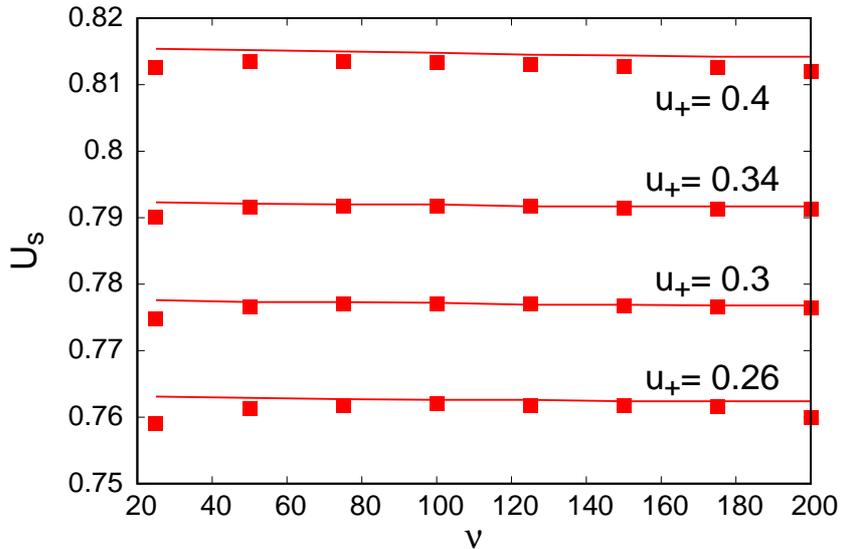}
\caption{Comparison between the Whitham shock velocity $U_{s}$ given by the modulation equation jump conditions (\ref{e:massjump})--(\ref{e:energyjump}) with (\ref{e:vrcm}) and numerical solutions of the nematic equations (\ref{e:eeqnd}) and (\ref{e:direqnd}) in the TDSW regime.  Solution of jump conditions:  red (full) line; numerical solution: red boxes.  Here $u_{-}=1.0$ and $q=2$.}
\label{f:whitham}
\end{figure}

The Whitham modulation equations in the highly nonlocal limit $\nu \gg 1$ (low optical power) have previously been determined \cite{saleh}, and so these modulation equations will just be quoted here.  These modulation equations determine the mean level $\bar{\rho}$ of $\rho$,
the amplitude $a$ and the wavenumber $k$ of the Stokes' wave solution of the nematic equations (\ref{e:mass})--(\ref{e:thm}).  As there is no known general periodic wave
solution of the nematic equations, the highly nonlocal Whitham modulation equations are derived based on the weakly nonlinear Stokes wavetrain for the nematic equations \cite{saleh}.  It can be seen from Figure \ref{f:types}(d)
that the resonant wavetrain has small amplitude, so the
weakly nonlinear limit is appropriate.
In the highly nonlocal limit $\nu \gg 1$, the Stokes' wave solution of the nematic equations (\ref{e:mass})--(\ref{e:thm}) is
\begin{eqnarray}
 \rho & = & \bar{\rho} + a \cos \varphi + \ldots
 \label{e:rhostokes} \\
 v & = & \bar{v} + av_{1}\cos \varphi + \ldots
 \label{e:vstokes} \\
 \phi & = & \frac{\bar{\rho}}{q} + a\phi_{1} \cos \varphi + \ldots
 \label{e:thetastokes} \\
 \omega & = & \omega_{0} + a\omega_{1} + a^{2} \omega_{2} + \ldots, \label{e:omstokes}
\end{eqnarray}
where the uniform phase is $\varphi = kx - \omega z$ and the over bar, $\bar{\rho}$ and $\bar{v}$, denotes the mean value of a wave parameter \cite{saleh}.  The amplitude $a$ of the (Stokes) wave is assumed to be small.   The work of \cite{saleh} gives that at $O(a)$ the nematic Stokes coefficients and the $O(a^{2})$ correction to the dispersion relation are
\begin{eqnarray}
\omega_{0} & = & k\bar{v} + \frac{k^2}{2} + \frac{4\bar{\rho}}{\nu k^2} - \frac{8\bar{\rho}q}{\nu^2k^4} - \frac{16\bar{\rho}^2}{\nu^2k^6} + \ldots, \label{omega0approx} \\
v_{1} & = & {\frac {k}{2\bar{\rho}}}+{\frac {4}{\nu\,{k}^{3}}}-{\frac {8q}{{
\nu}^{2}{k}^{5}}}-{\frac {16\bar{\rho}}{{\nu}^{2}{k}^{7}}} + \ldots, \label{v1approx} \\
\phi_{1} & = & \frac{2}{\nu k^2} - \frac{4q}{\nu^2k^4} + \ldots, \label{phi1approx} \\
\omega_{2} & = & -{\frac {{k}^{2}}{8\bar{\rho}^2}}-{\frac {3}{\bar{\rho}\nu k^2}}+{\frac {6q}{\bar{\rho}\nu^{2}k^{4}}}-{
\frac {20}{\nu^{2}k^6}} + \ldots \label{omega2approx}
\end{eqnarray}
The coefficient $\omega_{1}$ is set to zero, $\omega_{1}=0$, to eliminate secular terms. The weakly nonlinear Whitham modulation equations can then be derived by averaging conservation laws deduced from N\"{o}ther's Theorem \cite{gelfand}.  The Lagrangian for the hydrodynamic form of the nematic equations (\ref{e:mass})--(\ref{e:thm}) is
\begin{equation}
 L = -2\rho \psi_{z} - \frac{1}{4} \frac{\rho_{x}^{2}}{\rho} - \rho \psi_{x}^{2} - 4\rho \phi
 + \nu \phi_{x}^{2} + 2q\phi^{2}.
 \label{e:lag}
\end{equation}
Applying N\"other's Theorem we have that translation invariance with respect to the phase $\psi$ gives the mass conservation law
\begin{equation}\label{Lagrangmass}
    \frac{\partial}{\partial z}\frac{\partial L}{\partial \psi_{z}} + \frac{\partial}{\partial x}\frac{\partial L}{\partial \psi_{x}} = 0,
\end{equation}
translation invariance with respect to space $x$ yields the momentum conservation law
\begin{equation}\label{Lagrangmomentum}
    \frac{\partial}{\partial z} \left( \frac{\partial L}{\partial \rho_{z}} \frac{\partial \rho}{\partial x} + \frac{\partial L}{\partial \psi_{z}} \frac{\partial \psi}{\partial x} + \frac{\partial L}{\partial \phi_{z}} \frac{\partial \phi}{\partial x}\right) + \frac{\partial}{\partial x} \left( \frac{\partial L}{\partial \rho_{x}} \frac{\partial \rho}{\partial x} + \frac{\partial L}{\partial \psi_{x}} \frac{\partial \psi}{\partial x} + \frac{\partial L}{\partial \phi_{x}} \frac{\partial \phi}{\partial x} - L\right) = 0.
\end{equation}
and translation invariance with respect to time-like $z$ gives the energy conservation law
\begin{equation}\label{Lagrangenergy}
    \frac{\partial}{\partial z} \left( \frac{\partial L}{\partial \rho_{z}} \frac{\partial \rho}{\partial z} + \frac{\partial L}{\partial \psi_{z}} \frac{\partial \psi}{\partial z} + \frac{\partial L}{\partial \phi_{z}} \frac{\partial \phi}{\partial z} - L \right) + \frac{\partial}{\partial x} \left( \frac{\partial L}{\partial \rho_{x}} \frac{\partial \rho}{\partial z} + \frac{\partial L}{\partial \psi_{x}} \frac{\partial \psi}{\partial z} + \frac{\partial L}{\partial \phi_{x}} \frac{\partial \phi}{\partial z} \right) = 0.
\end{equation}
Substituting the Stokes expansions (\ref{e:rhostokes})--(\ref{e:thetastokes}) into these conservation laws and averaging by integrating in $\varphi$ from $0$ to $2\pi$ \cite{whitham} yields the modulation equations, truncated to $O(1/\nu)$,
\begin{eqnarray}
& & \frac{\partial k}{\partial z}+\frac{\partial}{\partial x}\left( k\bar{v} + \frac{k^2}{2} + \frac{2\bar{\rho}}{q} + \frac{4\bar{\rho}}{\nu k^2} -{\frac {{k}^{2}a^2}{8\bar{\rho}^2}} - \frac{3a^2}{\nu \bar{\rho}k^2} \right) = 0, \label{e:waveconv} \\
& & \frac{\partial \bar{\rho}}{\partial z} + \frac{\partial}{\partial x}\left( \bar{\rho}\bar{v} + \frac{ka^2}{4\bar{\rho}} + \frac{2a^2}{\nu k^3} \right) = 0, \label{e:massconv} \\
& &     \frac{\partial}{\partial z}\left( \bar{\rho}\bar{v} + \frac{ka^2}{4\bar{\rho}} + \frac{2a^2}{\nu k^3} \right) + \frac{\partial}{\partial x} \left( \frac{\bar{\rho}^2}{q} + \bar{\rho}\bar{v}^2 + \frac{k^2a^2}{4\bar{\rho}} + \frac{\bar{v}ka^2}{2\bar{\rho}} + \frac{4\bar{v}a^2}{\nu k^3} + \frac{a^2}{\nu k^2} \right) = 0, \label{e:momconv} \\
    & & \frac{\partial}{\partial z} \left( \bar{\rho}\bar{v}^2 + \frac{2\bar{\rho}^2}{q} + \frac{k^2a^2}{4\bar{\rho}} + \frac{k\bar{v}a^2}{2\bar{\rho}} + \frac{4a^2}{\nu k^2} + \frac{4\bar{v}a^2}{\nu k^3} \right) + \frac{\partial}{\partial x}\left( \bar{\rho}\bar{v}^3 + \frac{4\bar{v}\bar{\rho}^2}{q} + \frac{ka^2}{q} + \frac{k^3a^2}{4\bar{\rho}}  \right. \nonumber \\
        & & \left. \mbox{} + \frac{3\bar{v}k^2a^2}{4\bar{\rho}} + \frac{3\bar{v}^2ka^2}{4\bar{\rho}} + \frac{6\bar{v}a^2}{\nu k^2} + \frac{6\bar{v}^2a^2}{\nu k^3} + \frac{2a^2}{\nu k} \right )= 0 \label{e:energyconv}
\end{eqnarray}
for the (slowly varying) amplitude $a$, wavenumber $k$ and means $\bar{\rho}$ and $\bar{v}$ of the modulated Stokes wave \cite{whitham}.

The modulation equation (\ref{e:massconv}) is that for optical power conservation, equation (\ref{e:momconv}) is momentum conservation and (\ref{e:energyconv}) is energy conservation.  The modulation equation (\ref{e:waveconv}) is the equation for conservation of waves, $k_{z} + \omega_{x} = 0$, on noting that the $x$ derivative term is just the dispersion relation (\ref{e:omstokes}).   In this regard, it should be noted that the dispersion relation for the Stokes' wave from which the modulation equations are calculated has the carrier waves' phase shift term $2\bar{\rho}/q$ added \cite{nemgennady,saleh}, as explained above,
\begin{equation}
 \omega = k\bar{v} + \frac{1}{2}k^{2} + \frac{2\bar{\rho}}{q} + \frac{4\bar{\rho}}{\nu k^2} - \frac{k^{2}a^{2}}{8\bar{\rho}^{2}} - \frac{3a^{2}}{\nu k^{2}\bar{\rho}}.
 \label{e:stokeslarge}
\end{equation}

Figure \ref{f:types}(d) shows that in the TDSW regime the KdV-type nematic bore structure of Figures \ref{f:types}(a)--(c) has disappeared, leaving a dominant resonant wavetrain which is linked to the intermediate level $u_{i}$ by a negative polarity solitary wave \cite{patkdv}.  As discussed above, this link between the resonant wavetrain and the intermediate level can be approximated by a Whitham shock, a jump in the modulation equation variables \cite{saleh,patjump}, so that the Whitham shock links the resonant wavetrain with the level $u_{i}$ behind, in a similar manner to a gas dynamic shock wave links two compressible flow states \cite{whitham}.  Let us denote the amplitude, wavenumber, mean level and mean phase gradient of the resonant wavetrain by $a_{r}$, $k_{r}$, $\bar{\rho}_{r}$ and $\bar{v}_{r}$, respectively.  Matching the Whitham shock velocity $U_{s}$ to the Stokes' wave velocity (\ref{e:stokeslarge}), as these are co-propagating, gives
\begin{equation}
 U_{s} = \bar{v}_{r} + \frac{1}{2}k_{r} + \frac{2\bar{\rho}_{r}}{qk_{r}}
 + \frac{4\bar{\rho}_{r}}{\nu k_{r}^{3}} - \frac{k_{r}a_{r}^{2}}{8\bar{\rho}_{r}^{2}} - \frac{3a_{r}^{2}}{\nu k_{r}^{3} \bar{\rho}_{r}}.
 \label{e:resunon}
\end{equation}
Ahead of the Whitham shock there is the resonant wavetrain and behind it is a flat shelf, the intermediate level, which is a wavetrain of zero amplitude.  The mass, momentum and energy conservation equations (\ref{e:massconv})--(\ref{e:energyconv}) then give the jump conditions
\begin{eqnarray}
& &  U_{s} \left( \bar{\rho}_{r} - \rho_{i} \right) = \bar{\rho}_{r}\bar{v}_{r} + \frac{k_{r}a_{r}^2}{4\bar{\rho}_{r}} + \frac{2a_{r}^2}{\nu k^3_{r}} - \rho_{i}v_{i}, \label{e:massjump} \\
& & U_{s}\left( \bar{\rho}_{r}\bar{v}_{r} + \frac{k_{r}a_{r}^2}{4\bar{\rho}_{r}} + \frac{2a_{r}^2}{\nu k_{r}^3} - \rho_{i}v_{i} \right) = \frac{\bar{\rho}_{r}^2}{q} + \bar{\rho}_{r}\bar{v}_{r}^2 + \frac{k_{r}^2a_{r}^2}{4\bar{\rho}_{r}} + \frac{\bar{v}_{r}k_{r}a_{r}^2}{2\bar{\rho}_{r}} + \frac{4\bar{v}_{r}a_{r}^2}{\nu k_{r}^3} \nonumber \\
& & \mbox{} + \frac{a_{r}^2}{\nu k_{r}^2} - \frac{\rho_{i}^{2}}{q} - \rho_{i}v_{i}^{2}, \label{e:momjump} \\
& & U_{s} \left( \bar{\rho}_{r}\bar{v}_{r}^2 + \frac{2\bar{\rho}_{r}^2}{q} + \frac{k_{r}^2a_{r}^2}{4\bar{\rho}_{r}} + \frac{k_{r}\bar{v}_{r}a_{r}^2}{2\bar{\rho}_{r}} + \frac{4a_{r}^2}{\nu k_{r}^2} + \frac{4\bar{v}_{r}a_{r}^2}{\nu k_{r}^3} - \rho_{i}v_{i}^{2} - \frac{2 \rho_{i}^{2}}{q} \right) = \bar{\rho}_{r}\bar{v}_{r}^3 + \frac{4\bar{v}_{r}\bar{\rho}_{r}^2}{q} \nonumber \\
& & \mbox{} + \frac{k_{r}a_{r}^2}{q} + \frac{k_{r}^3a_{r}^2}{4\bar{\rho}} + \frac{3\bar{v}_{r}k_{r}^2a_{r}^2}{4\bar{\rho}_{r}} + \frac{3\bar{v}_{r}^2k_{r}a_{r}^2}{4\bar{\rho}_{r}} + \frac{6\bar{v}_{r}a_{r}^2}{\nu k_{r}^2} + \frac{6\bar{v}_{r}^2a_{r}^2}{\nu k_{r}^3} + \frac{2a_{r}^2}{\nu k_{r}} - \rho_{i}v_{i}^{3} - \frac{4v_{i}\rho_{i}^{2}}{q} \label{e:energyjump}
\end{eqnarray}
across the Whitham shock.
Together with the resonance condition (\ref{e:resunon}), these form four equations for the five unknowns $U_{s}$, $a_{r}$, $k_{r}$, $\bar{\rho}_{r}$ and $\bar{v}_{r}$, noting that $\rho_{i}$ and $v_{i}$ are given by (\ref{e:shelf}) and (\ref{e:v2expr}), respectively. The final equation is obtained by assuming that the Riemann invariant $R_{-}$ (\ref{e:cm}) is conserved through the resonant wavetrain and its lead partial DSW, which is valid for a full DSW \cite{elreview,chaos}.  This then determines the mean of the resonant phase gradient $\bar{v}_{r}$ \cite{saleh}.   This
Riemann invariant condition gives
\begin{equation}
 \bar{v}_{r} = 2\sqrt{\frac{2}{q}} \left( \sqrt{\bar{\rho}_{r}} - \sqrt{\rho_{+}} \right) =
 2\sqrt{\frac{2}{q}} \left( \sqrt{\bar{\rho}_{r}} - u_{+} \right).
 \label{e:vrcm}
\end{equation}

The above nematic Whitham modulation equation jump conditions (\ref{e:massjump})--(\ref{e:energyjump}) with (\ref{e:vrcm}) can be solved numerically for $U_{s}$, $a_{r}$, $k_{r}$ and $\bar{\rho}_{r}$ using Newton's method.  The full details for this numerical solution of the Whitham shock jump conditions can be found in \cite{saleh}.  Figure \ref{f:whitham} shows comparisons for the Whitham shock velocity from the nonlocal to local limits, the optical power increasing, as given by the jump conditions and full numerical solutions of the nematic equations (\ref{e:eeqnd}) and (\ref{e:direqnd}).
The values of the level ahead $u_{+}$ were chosen to lie in the TDSW regime,
see Figure \ref{f:regions}.
It can be seen that there is excellent agreement between the theoretical
and numerical solutions from high nonlocality, $\nu$ large, down to $\nu = O(10)$.  As for the lead solitary wave amplitude comparison of Figure \ref{f:pdswrdswamp} there is little change in the Whitham shock velocity
as the nonlocality parameter $\nu$ varies by an order of magnitude, with only a small increase in the velocity.  There is a small, increasing deviation between the theoretical and numerical values towards $\nu = 20$.   This is due to the onset on the VDSW regime for which $u$ vanishes at a point, a vacuum point \cite{nemgennady,saleh}.  Once the vacuum point is reached, $|u|$ cannot decrease further, so that the Whitham shock jump conditions need to be modified \cite{saleh}.  This will not be pursued further here.

For a fixed level ahead $u_{+}$, as the nonlocality parameter $\nu$ decreases (optical power increases),
the nematic DSW evolves from CDSW to TDSW type, see Figure \ref{f:regions}.
The borderline between the CDSW and TDSW regimes can be
determined from the Whitham shock jump conditions (\ref{e:massjump})--(\ref{e:energyjump}) and the resonance condition (\ref{e:resunon})
based on the following condition.  For a fixed nonlocality parameter $\nu$,
as the level ahead $u_{+}$ increases, it is found that the Whitham shock
velocity becomes greater than the linear group velocity
\begin{equation}
 c_{g} = \bar{v}_{r} + k_{r} - \frac{8\bar{\rho}_{r}}{\nu k_{r}^{3}}
 \label{e:lingroup}
\end{equation}
of the resonant wavetrain.  This is unphysical as this would mean that the
resonant wavetrain could not form.  Figure \ref{f:regions} shows this theoretical bound between the CDSW and TDSW regimes as a red line.  It can
be seen that the agreement with numerical solutions is excellent over the entire range of $\nu$, even for jump heights $u_{-} - u_{+}$ which are not
small.

As $\nu$ decreases, it would be expected that the high nonlocality modulation
equations (\ref{e:waveconv})--(\ref{e:energyconv}) cease to be applicable.
The Whitham modulation equations for the nematic equations in the local limit,
$\nu$ small, were calculated based on the equivalent of the Stokes' wave expansions (\ref{e:rhostokes})--(\ref{e:omstokes}) and (\ref{omega0approx})--(\ref{omega2approx}), expanding in $\nu$ rather than $1/\nu$.  However,
these were found not to give solutions in good agreement with numerical solutions.  This is expected as Figures \ref{f:regions} and \ref{f:pdswrdswamp} show that the nematic DSW is highly nonlocal down to
very small values of $\nu$, which are unphysical, as discussed at the end
of Section \ref{s:nonlocallocal}.

\section{RNLS and NLS DSW Regimes}


The KdV approximation (\ref{e:kdv5nem}) gives that the DSW changes from KdV to NLS type at $\nu = q^{2}/(4u_{+}^{2})$ due to the change in sign of the coefficient of the third derivative, as discussed above, and as shown in Figure \ref{f:regions}.
It should be noted that numerical solutions do not show a distinct change in DSW type, but a transition from a KdV-type DSW to an NLS-type DSW, as seen in Figures \ref{f:types}(d) and (e) and as shown by the two regimes RNLS and NLS type in Figure \ref{f:regions}.  The TDSW regime is characterised by a negative polarity solitary wave connecting the resonant wavetrain to the intermediate level, as in Figure \ref{f:types}(d).  As the nonlocality $\nu$ decreases, the beam power in increases, the height of this solitary wave decreases, resulting in the RNLS regime which
consists of a Whitham shock connecting a resonant wavetrain
to the intermediate level $u_{i}$.  Ahead of the resonant wavetrain is a partial NLS-type DSW which connects to the level ahead $u_{+}$, see Figure \ref{f:types}(e).  As $\nu$
decreases further, the resonant wavetrain shrinks and the partial DSW becomes a full NLS DSW with solitary waves at
its trailing edge and linear waves at its leading edge, resulting in an NLS DSW for sufficiently small $\nu$, see
Figure \ref{f:regions}(b).  In the limit the NLS DSW alone links the intermediate level to the level ahead.  As $\nu$
decreases in the RNLS regime, the waves at the trailing
edge of the partial DSW evolve from weakly nonlinear Stokes
waves to fully nonlinear periodic waves, then solitary waves in the NLS DSW regime.

The intermediate RNLS state, illustrated in Figure \ref{f:types}(e), consists of a resonant wavetrain with the height of the negative polarity solitary wave at the
Whitham shock negligible.  To compare the KdV-NLS DSW boundary $\nu = q^{2}/(4u_{+}^{2})$ with numerical solutions, the choice of the height of this solitary wave being $5\times 10^{-3}$ above $u_{i}$ was chosen for the onset of the RNLS regime in numerical solutions.  It can be seen that there is good agreement for this regime boundary for $u_{+}$ close to $u_{-}$, but there is increasing disagreement as $u_{+}$ decreases.  This is expected as the reductive nematic eKdV equation (\ref{Pofq}) was derived under the assumption that $|u_{-} - u_{+}|$ is small.

\section{Conclusions}

The structure of the nematic DSW (dispersive shock wave) solution of the defocussing nematic equations governing the propagation of an optical beam through a cell filled with nematic liquid crystals has been investigated using a combination of numerical solutions of the equations governing the beam, consisting of an NLS-type equation for the optical beam and an elliptic medium response equation, and solutions of the governing nematic equations using Whitham modulation theory and/or asymptotic solutions.
In contrast to previous work \cite{nembore,nemgennady,saleh}
the evolution of the DSW structure was studied as the power of the optical beam varied, from the experimental low power for which the nematic response is nonlocal to high power
for which the response is local.  As the beam power varies, it was found that the nematic DSW transitions between six regimes, four of which were studied in previous work
\cite{nembore,nemgennady,saleh}.  The two NLS-type DSWs do not exist in the low power regime studied in this previous work.  However, the experimental verification of these high power DSW types is questionable as the powers required for
their existence are unrealistic due to the possible excessive medium heating the high beam powers would cause.
Excellent agreement was found between numerical solutions and analytical solutions for the four physically relevant DSW regimes, the PDSW, RDSW, CDSW and TDSW regimes displayed in Figure \ref{f:types}(a)--(d).  In particular, the analytical theory gives good agreement for the boundaries between the existence regions for five of the DSW types as the nonlocality parameter $\nu$ varies, the exception being the boundary between the PDSW and RDSW regimes.  

It has been found that the details of the nematic DSW, for instance its lead wave amplitude and velocity and the amplitude of the associated resonant radiation, are well approximated by the nematic equations in the high nonlocality limit, the nonlocality parameter $\nu$ large, as the DSW transitions to the local limit.  This holds for the PDSW, RDSW, CDSW and TDSW regimes with the nonlocality parameter $\nu$ ranging from $O(100)$ to $O(1)$.  As the analysis of the nematic DSW is much easier in the high nonlocality limit based on asymptotic analyses with $\nu \gg 1$, this is an important result for future analysis of the nematic DSW in its various regimes and over its nonlocality range.

There are still a number of issues which could be addressed by the future work.  
An outstanding issue is the correct determination of resonance between the nematic DSW and diffractive radiation.  In contrast to the high nonlocality limit with $\nu$ large \cite{nembore,nemgennady,saleh}, as $\nu$ decreases from the high to the low nonlocality limit, the beam power increases,
the resonance condition used in previous work in the RDSW and CDSW regimes that the velocity of the lead solitary wave of the DSW matches the linear phase velocity of the resonant waves does not agree with numerical solutions.  The theoretical transition from the PDSW regime for high nonlocality to the RDSW regime due to the onset of resonance as $\nu$ decreases
occurs for $\nu=O(1)$, while numerical solutions show the
transition for $\nu=O(100)$ to $O(10)$.  As noted in previous work \cite{saleh}, and in contrast with other work on resonant DSWs, not only is the lead wave of the DSW in resonance with diffractive radiation, but the whole modulated periodic wave which forms the DSW is in resonance.  However, even this observation does not yield the correct resonance condition for the RDSW regime as the nonlocality parameter $\nu$ decreases below the highly nonlocal limit.   The correct resonance condition between the DSW and diffractive radiation requires further study.  This should be an important general issue for all resonant DSWs beyond the specific application to nematic liquid crystals.

\section*{Acknowledgement}

Saleh Baqer thanks the research sector of Kuwait University for a Research Initiation Grant (RIG) No. [ZS02/21] given during the preparation of the paper.


\end{document}